\documentclass[preprint,journal]{vgtc}            

\onlineid{0}

\ieeedoi{10.1109/TVCG.2026.3747530}

\vgtccategory{Research}

\title{Exploring Cross-Reality Transitions between Projections and Head-Mounted Displays for Immersive Digital Art}

\author{%
  \authororcid{Xiangpeng Fu}{0009-0003-0977-7879},
  \authororcid{Mads Haahr}{0000-0002-9273-6458}
}

\authorfooter{
  \item
    X. Fu and M. Haahr are with Trinity College Dublin, Ireland.
  	E-mail: \{fuxi\,$|$\,haahrm\}@tcd.ie\,.
    
}

\abstract{%
  Immersive exhibitions increasingly combine large-scale projection displays and mixed reality (MR) head-mounted displays (HMDs), yet their integration into a coherent experience remains underexplored, particularly in terms of how users perceive transitions across heterogeneous visualization environments. This paper investigates cross-reality (CR) object- and scene-level transitions between projection and an MR HMD through a hybrid immersive art installation spanning projection, augmented reality (AR), and virtual reality (VR). In a within-subjects study (N=24), we compared a calibrated condition with a bundled inconsistency condition that introduced noticeable cross-display inconsistencies in spatial alignment, visual appearance, and cross-device latency. Rather than using this contrast to establish perceptual thresholds, acceptable error bounds, or hardware targets, we used it as a controlled diagnostic probe to make transition-relevant disruptions more perceptible and discussable. Quantitative results confirmed that the manipulation produced a clear experiential contrast, with lower presence and higher workload under inconsistency. Interviews further revealed that different cues shaped experience through different mechanisms: spatial misalignment disrupted action--outcome predictability during precision interaction, appearance mismatch affected aesthetic coherence, and latency reduced perceived responsiveness. These effects also varied across asset types, with rigid static assets exposing inconsistencies more clearly, animated skeletal assets shifting attention toward motion impression, and particle effects often masking small discrepancies. These findings provide empirical insight into how users perceive projection--MR transitions and inform asset-aware design strategies for coherent hybrid immersive art experiences. We also release the core system as an open-source Unreal Engine plugin, \textit{HUICRSync}, to support future research and prototyping of projection--MR CR experiences.
  
}

\keywords{Mixed reality, hybrid user interfaces, cross reality, immersive digital art}

\teaser{
  \centering
  \includegraphics[width=0.245\linewidth]{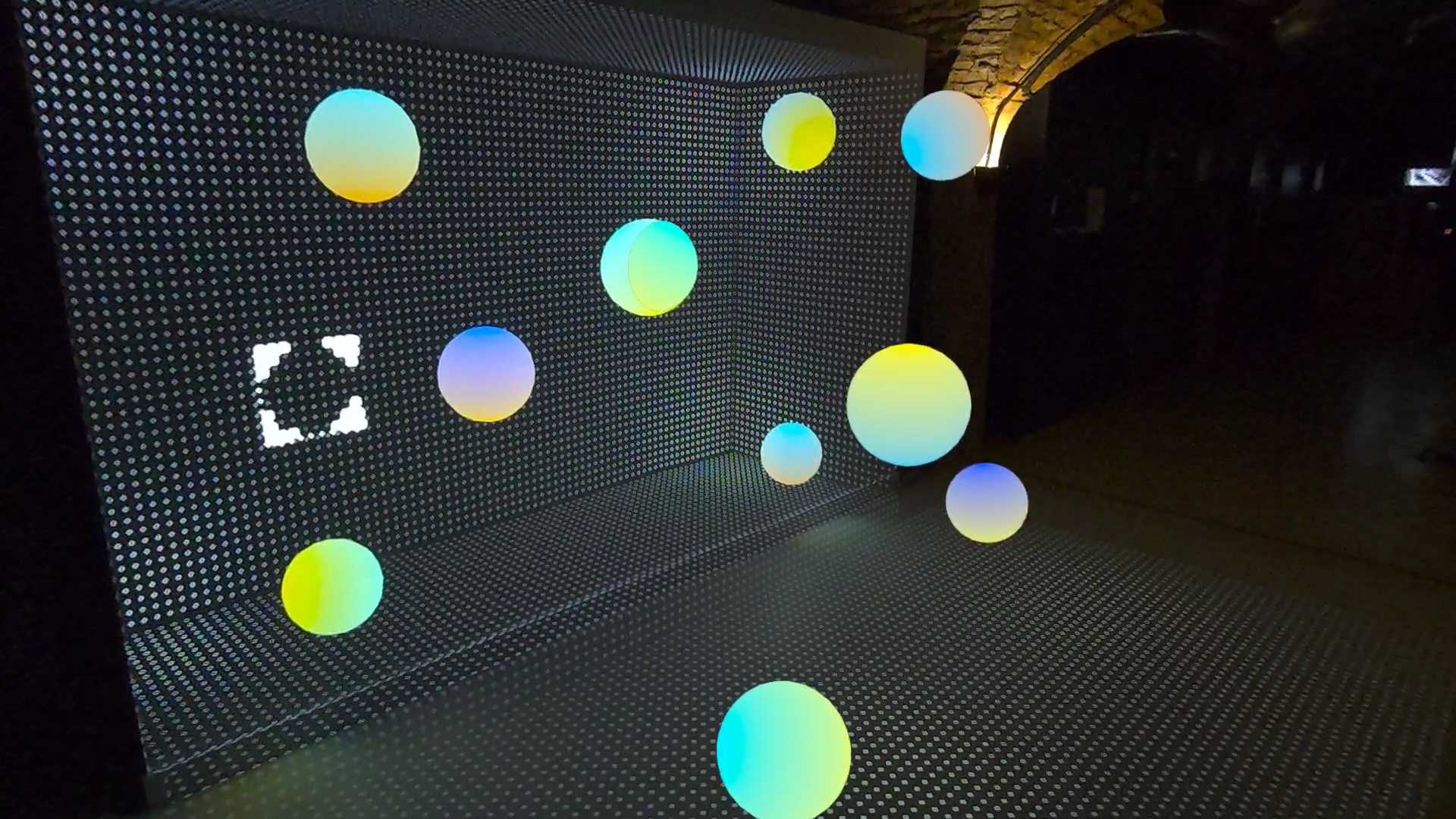}
  \hfill
  \includegraphics[width=0.245\linewidth]{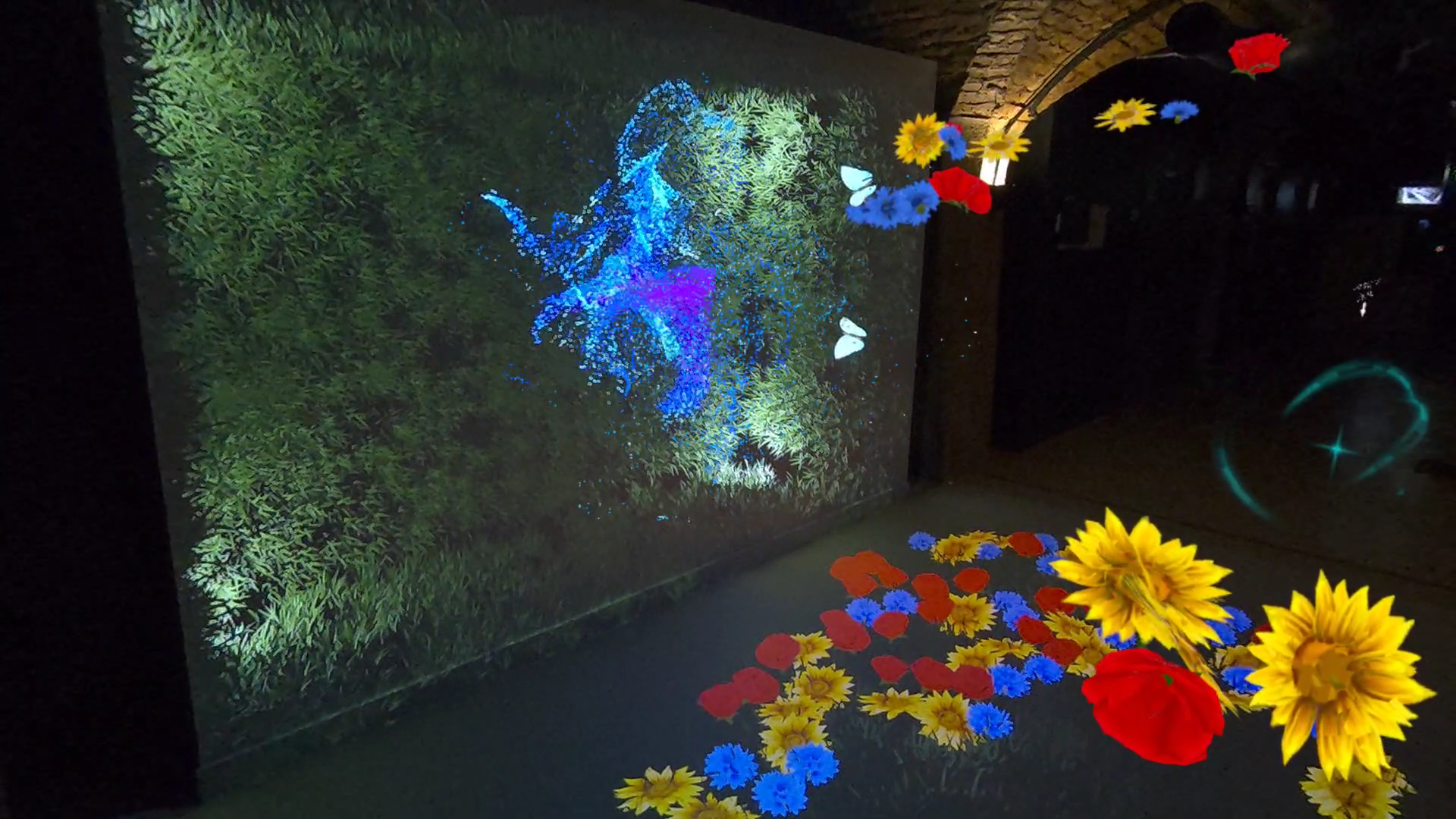}
  \hfill
  \includegraphics[width=0.245\linewidth]{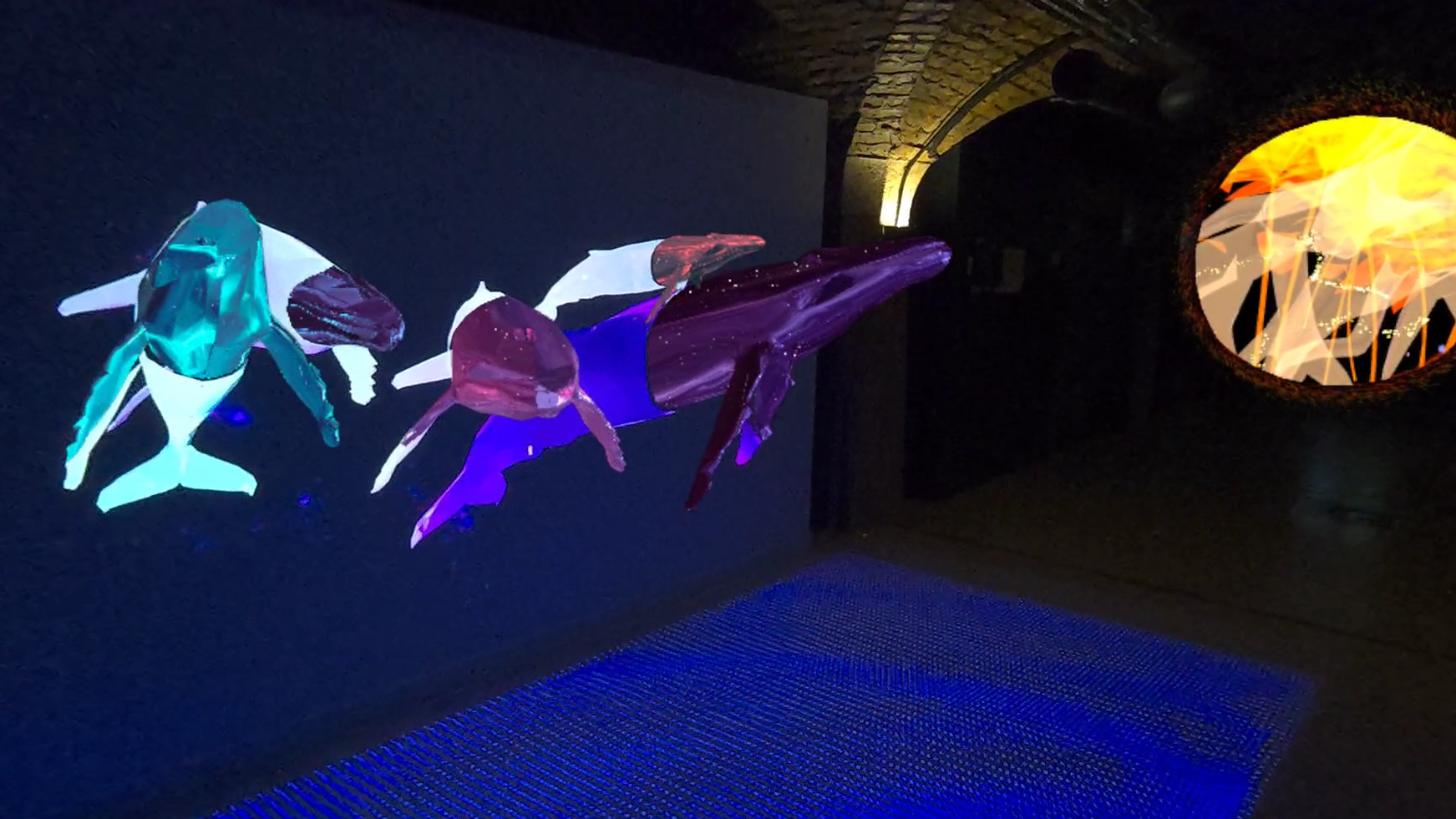}
  \hfill
  \includegraphics[width=0.245\linewidth]{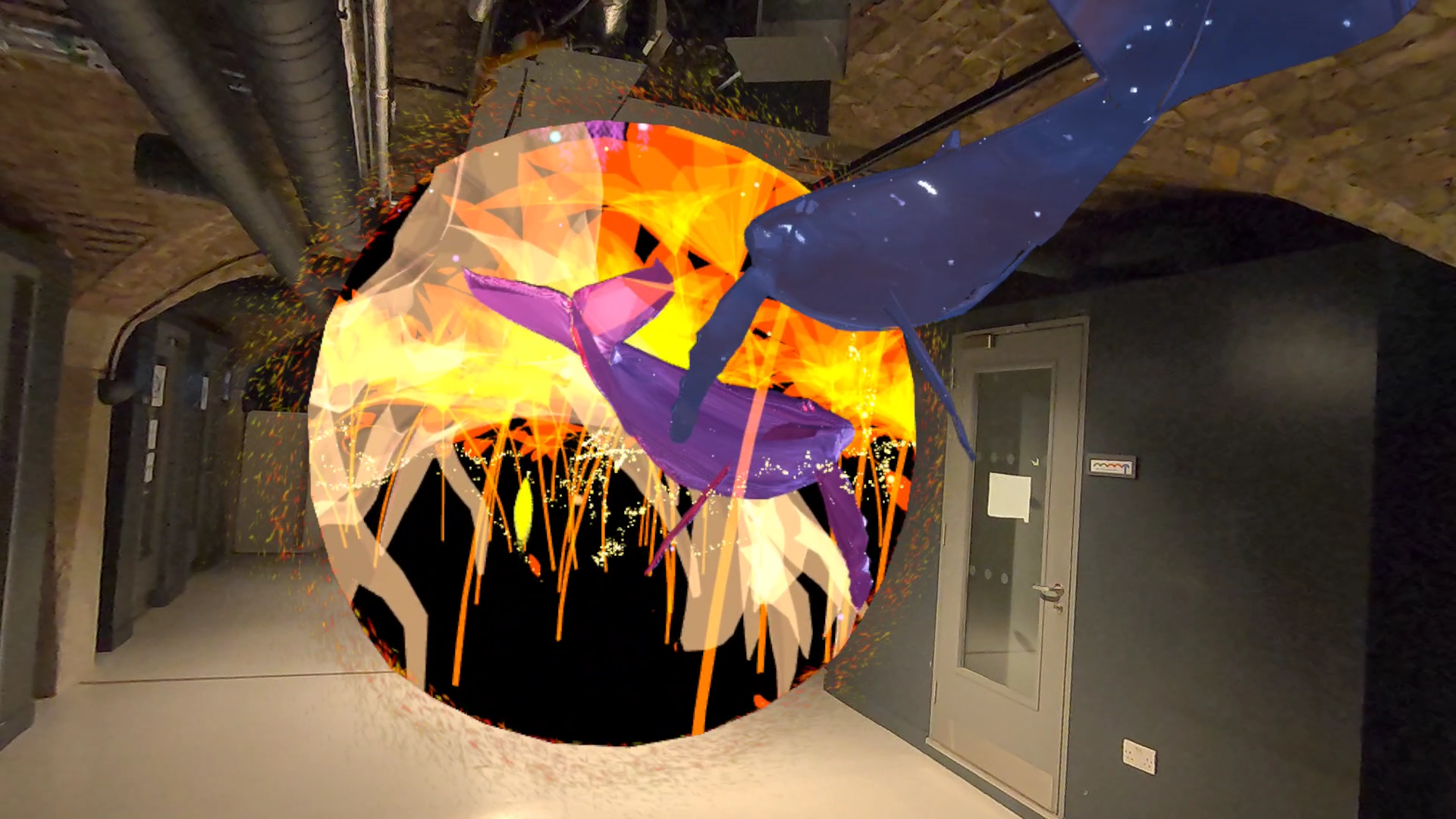}
  \caption{Photographs from four scenes of our hybrid projection--MR immersive art installation supporting cross-reality (CR) object- and scene-level transitions across projection, AR, and VR.}
  \label{fig:teaser}
}

\graphicspath{{figs/}{figures/}{pictures/}{images/}{./}} 

\usepackage{booktabs}                  
\usepackage{lipsum}                    
\usepackage{mwe}                       
\usepackage{ccicons}                   

\usepackage{amsmath}
\usepackage{graphicx}
\usepackage{enumitem}

\usepackage{mathptmx}                  

\begin{document}


\firstsection{Introduction}

\maketitle

Immersive exhibitions increasingly combine multiple physical and digital media to create spatial experiences \cite{Yu23}. In practice, large-scale projection often remains the central architectural and expressive medium, while headset-based experiences provide individualized viewpoints. These media are often presented side by side rather than as a coherent whole; for example, \emph{Van Gogh: The Immersive Experience} combines a large-scale projection environment with a separate VR component \cite{vangogh26}. This coexistence makes immersive digital art a compelling setting for studying hybrid experiences in which users move between different realities while engaging with artworks. Yet we still lack empirical understanding of how users perceive transitions between these heterogeneous visualization environments in such settings, and therefore lack clear guidance for designing them as coherent hybrid experiences.

Prior work has framed shifts of content and experience across devices as \emph{Cross-Reality} (CR) transitions along the Reality--Virtuality Continuum (RVC) \cite{Milgram94,Liang23}. In hybrid experiences, such transitions can disrupt continuity and introduce cognitive and attentional costs \cite{Zagermann22}. This is particularly relevant for projection--MR experiences, where content must remain perceptually coherent as it moves between physically and visually distinct media. In immersive art, where narrative coherence and sustained presence are central goals, such inconsistencies may be especially disruptive.

Although simulating 2D screens inside HMDs is one possible solution, our focus is different. In immersive art installations, physical projection and HMD content often already coexist, and projection is part of the spatial and aesthetic structure of the experience rather than a display surface to be replaced. We therefore focus on transitions that bridge projection and MR as distinct visualization environments. 

In this context, we treat visual continuity as one key analytical lens for examining transition coherence. This framing builds on the object- and environment-transition literature discussed in \cref{sec:related-work}. With this lens, our goals are twofold: first, to better understand how users perceive projection--MR transitions across heterogeneous visualization environments; and second, to use that understanding to inform the design of future hybrid immersive experiences. To pursue these goals, we structure our investigation progressively: identifying the perceptual cues involved in projection--MR transitions, examining their experiential consequences, and assessing whether these effects differ across common digital art asset types. This progression motivates the following research questions:

\begin{itemize}[noitemsep, topsep=0pt]
  \item \textbf{RQ1:} Which cues shape perceived transition quality?
  \item \textbf{RQ2:} How do these cues affect users' immersive experience?
  \item \textbf{RQ3:} Do different types of digital art assets show different sensitivity to these cues?
\end{itemize}

To answer these questions, we built a hybrid immersive art installation integrating large-scale projection, AR, and VR, and used it as an experimental platform in a controlled mixed-methods study ($N=24$). We compare a calibrated condition designed for coherent projection--MR transitions with a deliberately degraded condition that introduces noticeable cross-display inconsistencies. Rather than using this contrast to establish perceptual thresholds, acceptable error bounds, or hardware performance targets, we treat the deliberately degraded condition as a controlled diagnostic probe for design. By making transition disruptions more noticeable and discussable, this contrast allows us to examine which cues participants attend to and how these cues relate to perceived continuity, presence, workload, and asset-dependent experience. This paper contributes:
\begin{itemize}[noitemsep, topsep=0pt]
  \item a hybrid projection–MR immersive art installation for studying object- and scene-level CR transitions across projection, AR, and VR;
  \item mixed-methods findings on how spatial, appearance, and temporal cues shape perceived transition quality and immersive experience, including how these effects vary across different digital art asset types; and
  \item design implications for coherent hybrid projection--MR immersive art experiences.
\end{itemize}

In addition, we release the core system as an open-source Unreal Engine plugin, \textit{HUICRSync}, to support future research and prototyping of projection--MR CR experiences.

\section{Related Work}
\label{sec:related-work}
Our work builds on research on hybrid user interfaces (HUIs), which combine heterogeneous displays and interaction technologies within one experience \cite{Feiner91,Satkowski23}, and on CR transitions across heterogeneous visualization environments. We first review HUI application contexts, then focus on visual CR transitions at the object and environment levels to motivate our study of transition coherence in immersive digital art.

\subsection{Application Contexts of Existing HUI Systems}
HUIs combine heterogeneous displays and interaction technologies to leverage complementary strengths within a unified experience \cite{Feiner91,Satkowski23}. Although the term has broadened over time \cite{Satkowski23}, the idea of integrating MR HMDs with more traditional displays to support different tasks and perspectives remains consistent with its original intent.

Although many HUI systems do not explicitly frame their designs as CR transitions, they establish hybrid contexts in which users move across devices and immersion levels. A substantial body of work has explored such configurations in immersive analytics, visualization, and expert workflows, combining large displays, personal devices, and MR HMDs to support complementary immersive, spatial, and personal views \cite{Butscher18,Cavallo19,Reipschlager19,Reipschlager21,Langner21,Zhao25}. While these systems demonstrate how heterogeneous displays can be functionally coordinated, they less often examine how users perceive continuity as they move between display modalities.

By contrast, when transitions between realities are treated as an explicit object of study, the broader CR literature presents a different emphasis. A systematic literature review \cite{Pavavimol25} describes XR transition research as fragmented and often detached from concrete application contexts, with many studies focusing on general frameworks or techniques rather than practical use cases. Among work that does define a domain, games and entertainment dominate, with less attention to work, collaboration, and experiential or cultural settings. This discrepancy between existing hybrid practices and how CR transitions are studied motivates empirical investigation in experience-driven environments such as immersive digital art.

\subsection{Visual and Perceptual Transitions Across Reality Modalities}

We focus on visual and perceptual CR transitions, particularly visual continuity. While CR transitions have also been studied from interactional and system-level perspectives \cite{Rau25, Cools22, Cools25, Cools25_DP, Wang22}, we restrict our scope to how transitions are visually designed and perceived. Here, visual continuity concerns not only whether a transition appears smooth, but also whether users can continue to recognize object identity, spatial relation, and scene context as content crosses display boundaries. We organize prior work by the primary transitioning entity: objects and environments.

\subsubsection{Object Transitions}

Prior studies have shown that continuity-preserving object transitions can help users follow virtual objects, maintain mental mappings, and continue tasks across display spaces. Fischer et al. examine virtual object transitions between projector displays and AR glasses, showing that transition method, object size, and transition-path visibility affect immersion, discomfort, and ease of following the object~\cite{Fischer23}. In visualization contexts, Liao et al. and Schwajda et al. show that continuous transformations between immersive/non-immersive or 2D/3D representations can reduce cognitive effort and support task continuation~\cite{Liao25,Schwajda23}. Design-space and prototyping work further identifies geometric pose, surface relation, transformation control, and input modality as important dimensions for 2D/3D transformations and single-user CR object transitions~\cite{Lee22,Wang22_2}. More recent desktop--AR studies evaluate concrete techniques for moving 3D objects between monitor and AR spaces, including interaction-modality comparisons and domain-workflow evaluations~\cite{Cools25,Rau25}.

However, existing studies predominantly focus on specific object types or narrowly scoped scenarios, such as data visualizations or individual virtual artifacts. As a result, it remains unclear which transition factors are most critical for maintaining perceived continuity across diverse object categories, or how different asset types respond to transition inconsistencies. This is especially relevant in immersive digital art, where content spans structurally different forms such as static objects, animated skeletal characters, and particle-based effects, each of which may expose continuity disruptions differently. Accordingly, different asset types may require different transition design strategies.

\subsubsection{Environment Transitions}

Beyond object-level transitions, prior work has investigated transitions at the level of entire environments or scenes, including changes between virtual environments, between real and virtual environments, and across different stages of the RVC~\cite{Feld24,Husung19,Pointecker22,Pointecker24}. These studies examine common techniques such as cuts, fades, dissolves, portals, rifts, and object-based environmental transformations. Recent work highlights the context dependence of environment transitions: in task-driven CR settings with frequent switching, users often prefer brief and efficient transitions, whereas more interactive or visible techniques such as portals may be more appropriate for infrequent transitions, substantially different environments, or experiences that prioritize presence, continuity, and user acceptance~\cite{Husung19,Pointecker22,Feld24}. This context dependence is especially relevant for immersive digital art, where transitions are not only functional mechanisms for changing scenes but also part of the experiential and narrative staging of the work.

Portal-based transitions are an important class of environment transition because they provide a visible link to the destination environment: users can preview the target environment through the portal before physically traversing it~\cite{Feld24}. Recent work on VR world switching further shows that such portal previews support fast first-person pre-orientation before entry~\cite{Gottsacker26}. This makes portals particularly relevant to immersive exhibition settings, where visitors navigate through space and naturally move between artworks or scenes. In our work, this informed the use of an AR--VR portal in which users and animated assets traverse between environments through a spatially legible boundary.

Environment transitions also raise cognitive and perceptual challenges beyond the visual effect itself. Recent work on spatial cognitive residue shows that users may carry spatial memory and attentional resources from one virtual environment into the next, and that transition duration and intermediate visual content can influence how strongly users disengage from the previous environment~\cite{Gottsacker24}. Other qualitative work on CR transitions identifies common frictions such as disorientation, fear of the unknown, and physical or cognitive effort, and argues that transitions should help establish and preserve users' spatial mental models across realities~\cite{vonWillich25}. These findings motivated our use of gradual and spatially anchored scene transitions: between projection--AR scenes, a localized expansion effect gradually replaces the current scene and reveals the next while preserving spatial continuity; between AR and VR, a portal provides preview and active traversal.

Taken together, prior work provides important foundations for designing CR environment transitions, but object- and environment-level transitions are still often evaluated separately and outside experience-driven hybrid projection--MR settings. As a result, we still know little about how scene-level transition design and object-level cross-display consistency jointly shape perceived continuity in integrated projection--MR experiences. Our work addresses this gap through an empirical investigation in immersive digital art, with the aim of informing the design of coherent hybrid projection--MR experiences.

\section{Transition Cues and Asset Dimensions}

We structure the study around transition cues (RQ1-RQ2) and asset-related dimensions (RQ3).

\subsection{Transition Cues}
\label{subsec:Factors}
To identify transition-relevant cues for RQ1, we draw selectively on Jeong et al.'s affordance framework for interactive media art \cite{Rhee13}, particularly its cognitive and feedback-oriented aspects, to organize the cues that shape how users interpret continuity during CR transitions. In our context, cognitive aspects relate to whether content remains spatially and visually consistent across projection and MR, while feedback aspects relate to temporal responsiveness across devices.

Accordingly, we focus on three cue dimensions that are especially relevant to how users interpret continuity across media:

\begin{itemize}[noitemsep, topsep=0pt]
  \item \textbf{spatial alignment:} positional, rotational, and scale consistency across projection and the HMD.
  \item \textbf{visual appearance:}  color rendering consistency.
  \item \textbf{temporal responsiveness:} cross-device latency.
\end{itemize}

We operationalize the degraded condition by introducing controlled discrepancies along these three dimensions, and examine how they affect perceived transition quality, presence, and workload.

\subsection{Asset Types}
Asset structure and dynamics may mediate the perceptual impact of transition inconsistencies (RQ3). We therefore examine three common immersive-art asset types with distinct visual and behavioral properties:

\begin{itemize}[noitemsep, topsep=0pt]
  \item \textbf{Static meshes:} rigid, geometry-dominated objects with stable structure.
  \item \textbf{Skeletal meshes:} articulated objects whose appearance changes through animation and deformation.
  \item \textbf{Particle systems:} distributed, dynamic effects composed of many small elements.
\end{itemize}

\section{System and Transition Design}
\subsection{Hybrid System Overview}

Our study uses a hybrid projection--MR system that enables spatially coherent rendering and interaction across large-scale projection mapping and an MR HMD. The system consists of (1) a projection component rendering content onto physical surfaces, (2) an MR component rendering registered 3D content in the HMD, and (3) a calibration--synchronization layer that aligns coordinate frames and maintains consistent state across components during CR transitions. In this paper, the system serves as a fixed experimental platform. Pointers to the open-source implementation of this system, its documentation, accompanying example project, and archived version used in this work are provided in \hyperref[sec:supplemental_materials]{Supplemental Material Pointers}.

\subsection{Technical Setup}
\subsubsection{Devices and Implementations}
\label{subsubsec:system}
Two ceiling-mounted projectors create a wall and a floor projection surface (wall: $1920 \times 1200$, $3.45 \times 2.17\,\mathrm{m}$; floor: $1920 \times 1080$, $4.2 \times 2.37\,\mathrm{m}$), driven by a PC (i9-14900HX, RTX 4090 Laptop, 64 GB RAM). For MR, we use a standalone Meta Quest 3. Both projection and HMD applications are implemented in Unreal Engine 5 (C++) as separate instances. We synchronize object states, interaction events, and calibration data via a custom UDP client--server layer, which remains fixed throughout the study.

\subsubsection{Spatial Calibration}
\label{subsubsec:calibration}
\begin{figure}[t]
  \centering
  \begin{subfigure}[t]{0.46\linewidth}
    \centering
    \includegraphics[width=\columnwidth]{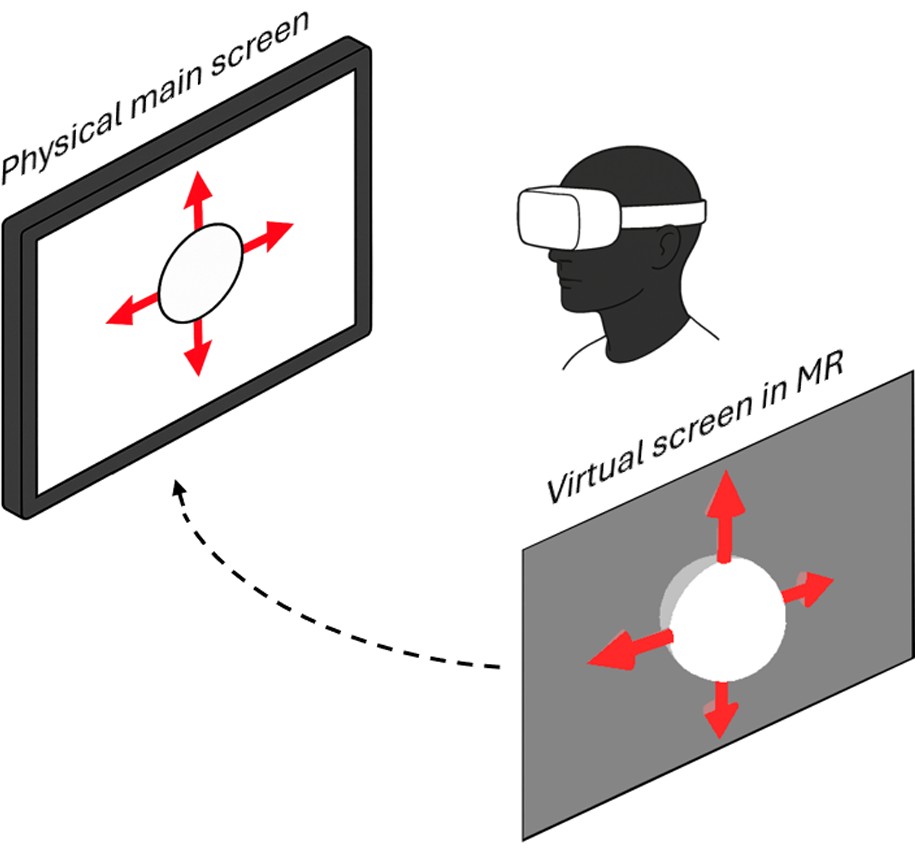}
    \caption{}
    \label{fig:SingleScreenCalibration1}
  \end{subfigure}
  \hfill
  \begin{subfigure}[t]{0.48\linewidth}
    \centering 
    \includegraphics[width=\columnwidth]{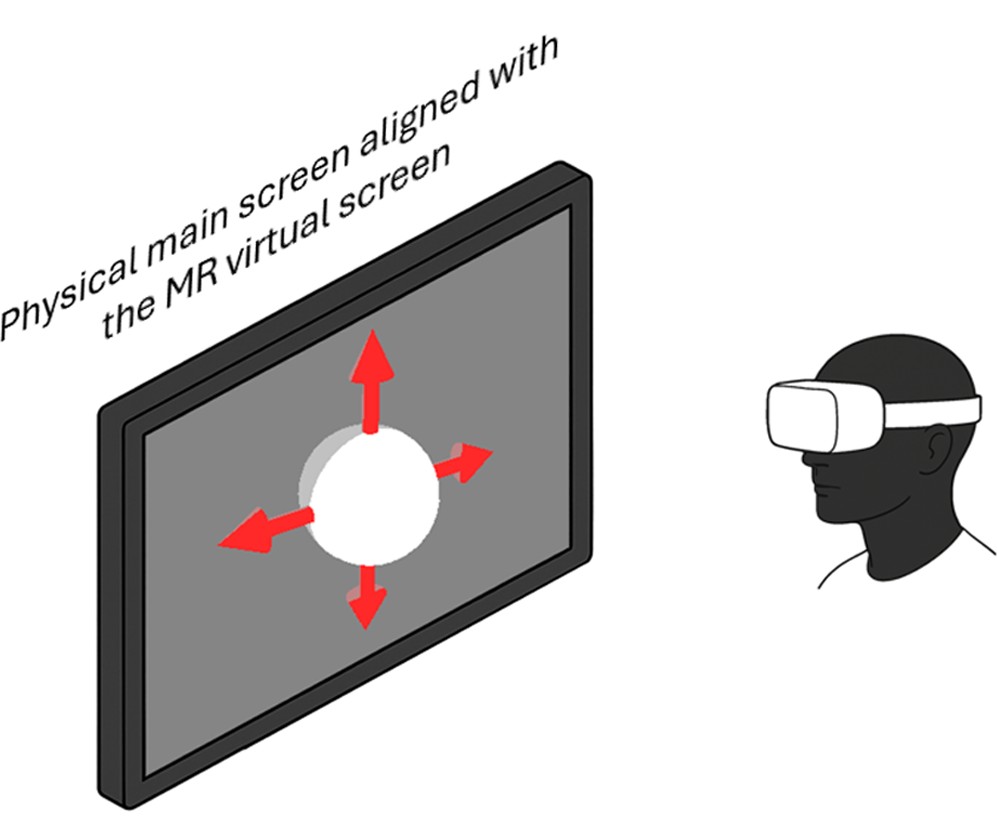}
    \caption{}
    \label{fig:SingleScreenCalibration2}
  \end{subfigure}
  \caption{Calibration of the MR HMD to the primary 2D display. (a) The virtual screen in MR is adjusted to align with the physical main screen. (b) Calibration is completed when the calibrator visually overlaps with its counterpart rendered on the physical display.}
  \label{fig:SingleScreenCalibration}
\end{figure}

To support coherent projection--MR transitions, we implemented spatial calibration between the projection surface and the MR headset. Similar spatial alignment strategies are common in hybrid display and projection-based systems, where they are used to establish geometric correspondence between physical and virtual display spaces and to support viewpoint-dependent rendering \cite{Reipschlager19, KAVE18, Zhao25}. In our setup, we establish a shared hybrid coordinate space by aligning each physical projection surface with an identical planar proxy in the MR view. We designate the wall projection as the primary screen and interactively adjust its MR proxy (position/orientation/scale) until a reference calibrator visually overlaps across modalities (\cref{fig:SingleScreenCalibration}). We record the transform of the main reference calibrator in the projection and HMD coordinate frames, denoted as $\mathbf{T}_{\mathrm{PC}}$ and $\mathbf{T}_{\mathrm{HMD}}$, respectively, and use them to compute the relative transformation:
\begin{equation}
\mathbf{T}_{\mathrm{PC}\rightarrow\mathrm{HMD}} 
= \mathbf{T}_{\mathrm{PC}}^{-1} \cdot \mathbf{T}_{\mathrm{HMD}}, \quad
\mathbf{T}_{\mathrm{HMD}\rightarrow\mathrm{PC}} 
= \mathbf{T}_{\mathrm{HMD}}^{-1} \cdot \mathbf{T}_{\mathrm{PC}}
\end{equation}

\begin{figure}[t]
  \centering
  \begin{subfigure}[t]{0.45\linewidth}
    \centering
    \includegraphics[width=\columnwidth]{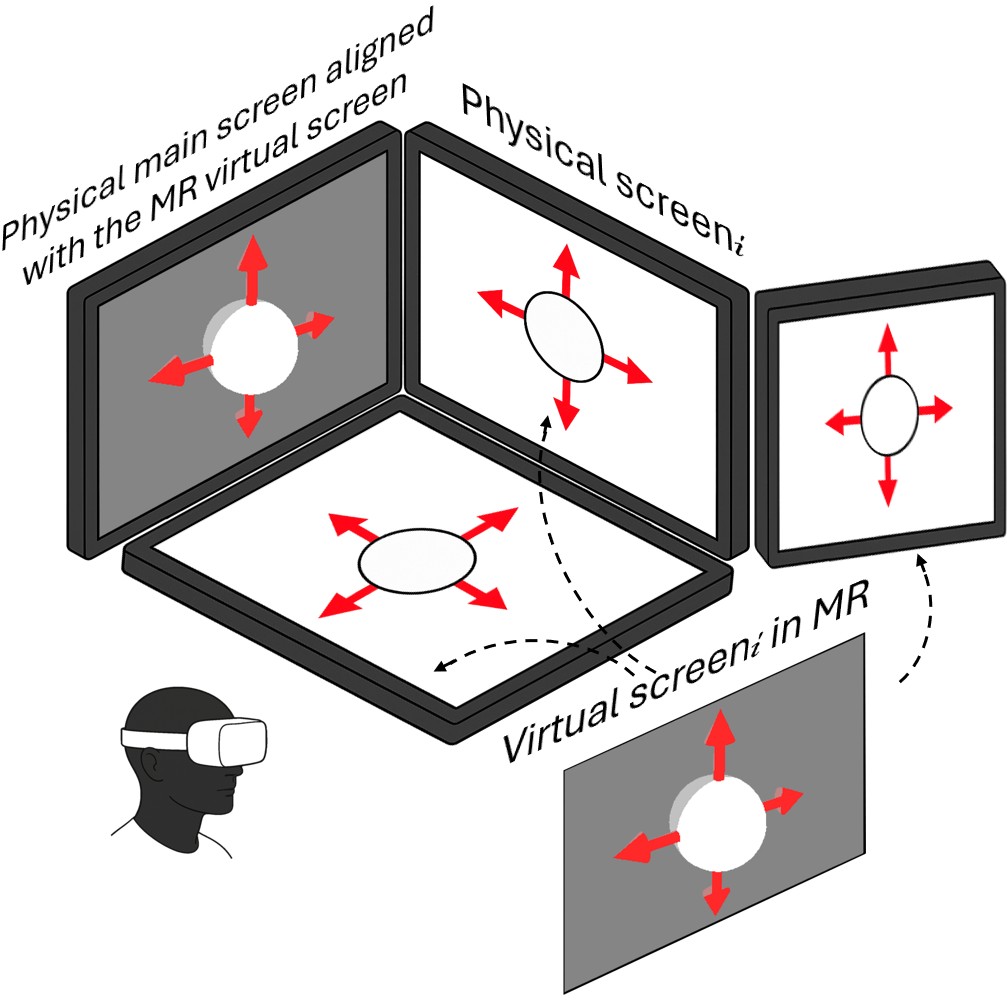}
    \caption{}
    \label{fig:MultipleScreenCalibration1}
  \end{subfigure}
  \hfill
  \begin{subfigure}[t]{0.45\linewidth}
    \centering 
    \includegraphics[width=\columnwidth]{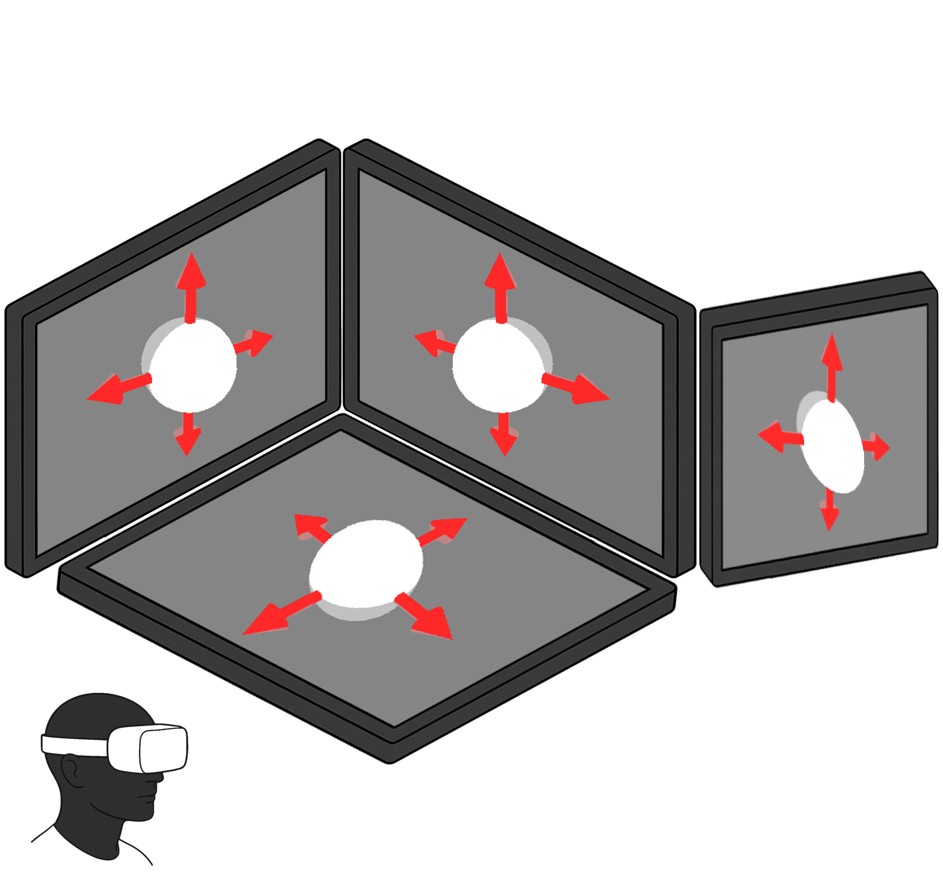}
    \caption{}
    \label{fig:MultipleScreenCalibration2}
  \end{subfigure}
  \caption{\textbf{(a)} Extending the calibration to additional screens. Each new screen (Screen$_i$) is visually aligned using the same calibrator-based calibration strategy.  \textbf{(b)} Final configuration with all 2D screens spatially aligned to their virtual representations in MR.}
  \label{fig:MultiScreenCalibration}
\end{figure}

After calibrating the primary screen, additional projection surfaces are registered using the same procedure (\cref{fig:MultiScreenCalibration}) and expressed in the projection coordinate system via \begin{equation}
\mathbf{T}_{\mathrm{screen}_i}^{\mathrm{(PC)}} =
\mathbf{T}_{\mathrm{screen}_i}^{\mathrm{(HMD)}} \cdot
\mathbf{T}_{\mathrm{HMD}\rightarrow\mathrm{PC}}
\end{equation}

This allows multiple 2D displays with arbitrary size, orientation, and spatial arrangement to be integrated into a unified hybrid coordinate space.

The resulting mappings are then used to transform object and camera poses consistently between the projection and MR applications during runtime via:

\begin{equation}
\mathbf{T}_{\mathrm{obj}}^{\mathrm{(HMD)}} =
\mathbf{T}_{\mathrm{obj}}^{\mathrm{(PC)}} \cdot
\mathbf{T}_{\mathrm{PC}\rightarrow\mathrm{HMD}}, \quad
\mathbf{T}_{\mathrm{obj}}^{\mathrm{(PC)}} =
\mathbf{T}_{\mathrm{obj}}^{\mathrm{(HMD)}} \cdot
\mathbf{T}_{\mathrm{HMD}\rightarrow\mathrm{PC}}
\end{equation}

\subsection{Transition Design}
\begin{figure}[t]
  \centering
  \begin{subfigure}[t]{0.25\columnwidth}
    \centering
    \includegraphics[width=\columnwidth]{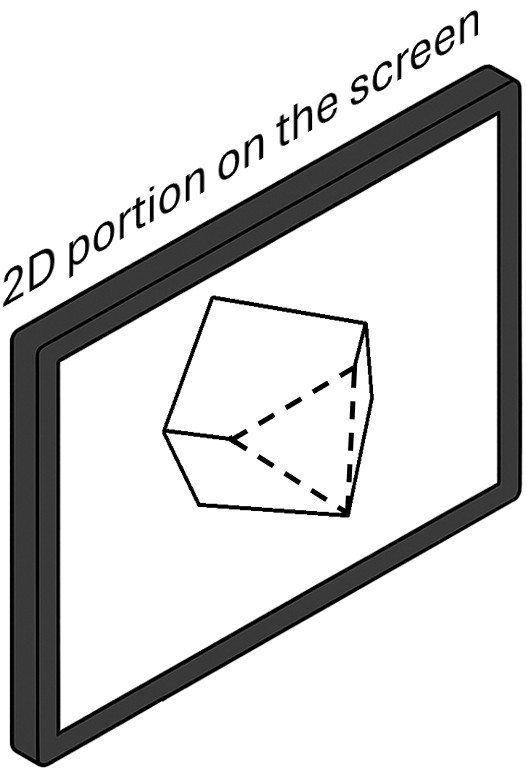}
    \caption{}
    \label{fig:Dissolve1}
  \end{subfigure}
  \hfill
  \begin{subfigure}[t]{0.25\linewidth}
    \centering 
    \includegraphics[width=\columnwidth]{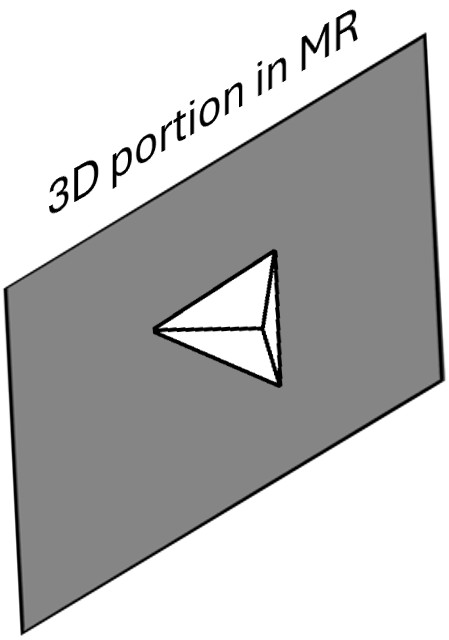}
    \caption{}
    \label{fig:Dissolve2}
  \end{subfigure}
  \hfill
  \begin{subfigure}[t]{0.4\linewidth}
    \centering 
    \includegraphics[width=\columnwidth]{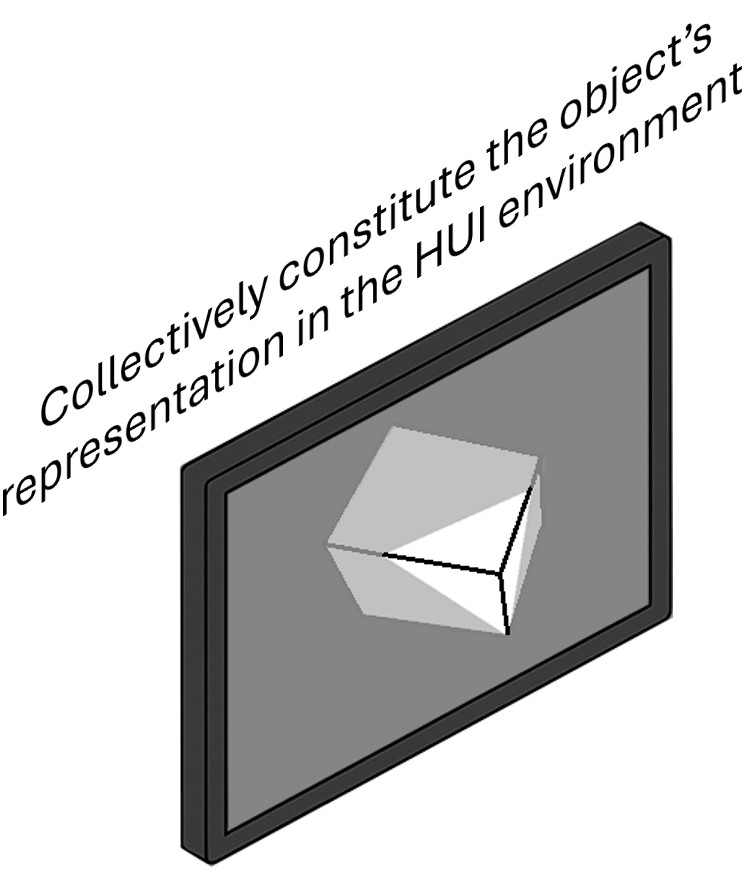}
    \caption{}
    \label{fig:Dissolve3}
  \end{subfigure}
  \caption{ \textbf{(a)} The 2D projection portion appears on the 2D display.  \textbf{(b)} The 3D portion remains in the MR HMD. \textbf{(c)} The 2D-rendered portion on the 2D display and the 3D-rendered portion in the MR environment collectively constitute the object’s representation in the hybrid space.}
  \label{fig:Dissolve}
\end{figure}

We realize visually continuous object transitions by spatially partitioning rendering across projection and MR. When an object intersects a projection surface, the portion inside the screen boundary is rendered only on the projection, while the remaining portion is rendered only in MR (\cref{fig:Dissolve}). When calibration is consistent, the two partial renderings are intended to be perceptually integrated as a single coherent object spanning both modalities.

Transitions are triggered using standard controller interactions \cite{Rau25}: users select targets via ray-based pointing and button input, and can directly grab/push objects at close range (including pushing toward a surface to return objects to projection).

\subsection{Baseline Appearance Tuning}
Before the study, we tuned both the PC-side projected content and the projector image settings to approximate the appearance of the corresponding MR-rendered content as viewed through the Quest~3 passthrough. This tuning was performed at both the asset and projector levels, adjusting appearance-related parameters where applicable, such as vividness, contrast, saturation, brightness, hue, and color tone, under the room-lighting conditions used in the study.

\section{Immersive Art Application for CR Evaluation}
\label{sec:Application}
We built an immersive art application on top of our hybrid projection--MR system to study CR transitions in an experience-driven setting. \Cref{fig:teaser} shows representative photographs from four scenes, spanning projection--AR interaction and an AR-to-VR portal transition. We describe the scene structure, interaction design, and how we operationalize the experimental conditions.

\subsection{Immersive Art Application Design}
\subsubsection{Experience Structure and Scenes}
\begin{figure*}[t]
  \centering
  \begin{subfigure}[t]{0.252\textwidth}
    \centering
    \includegraphics[width=\linewidth]{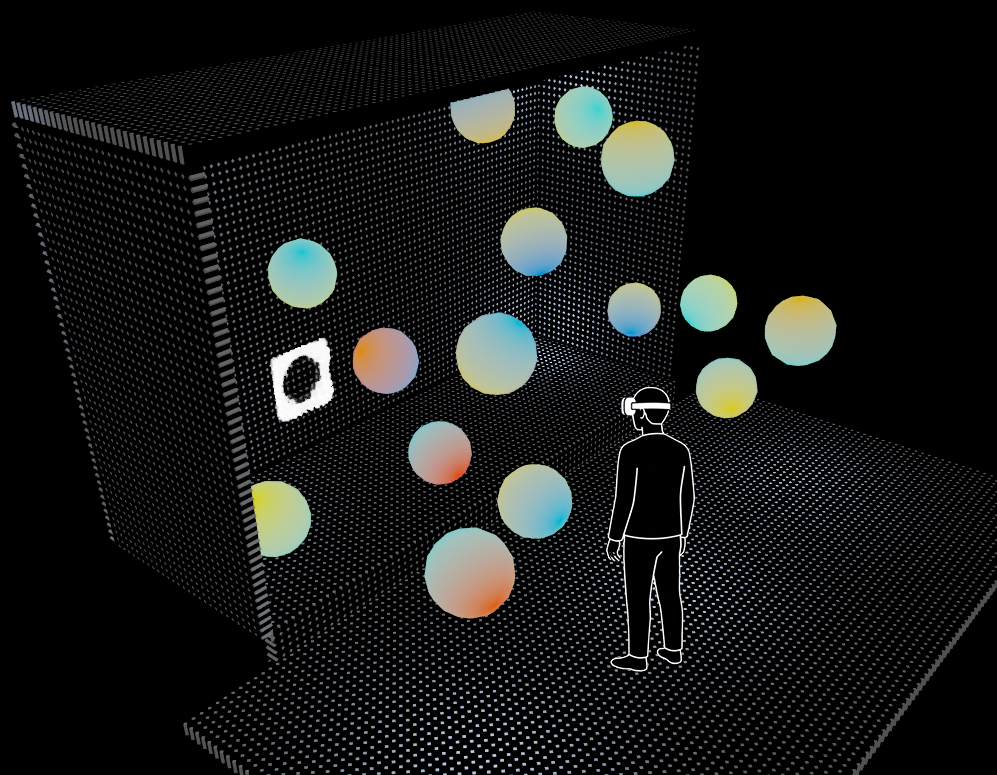}
    \caption{}
    \label{fig:scene1}
  \end{subfigure}
  \hfill
  \begin{subfigure}[t]{0.24\textwidth}
    \centering
    \includegraphics[width=\linewidth]{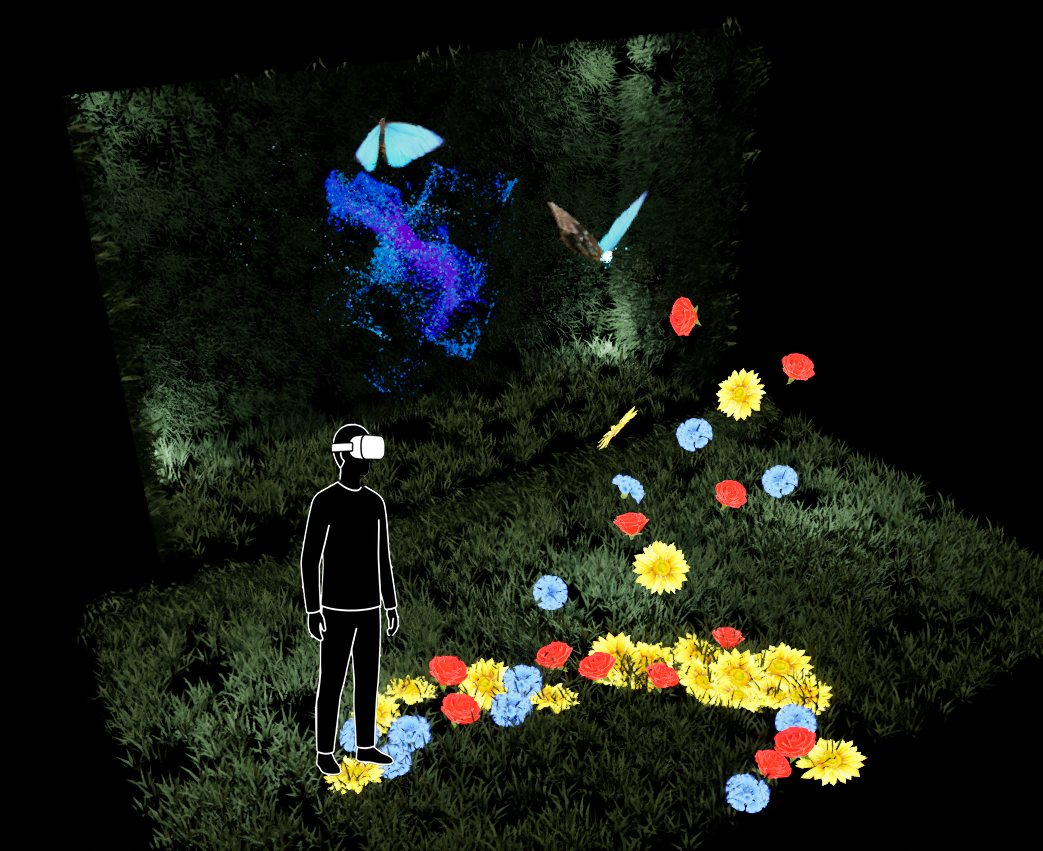}
    \caption{}
    \label{fig:scene2}
  \end{subfigure}
  \hfill
  \begin{subfigure}[t]{0.235\textwidth}
    \centering
    \includegraphics[width=\linewidth]{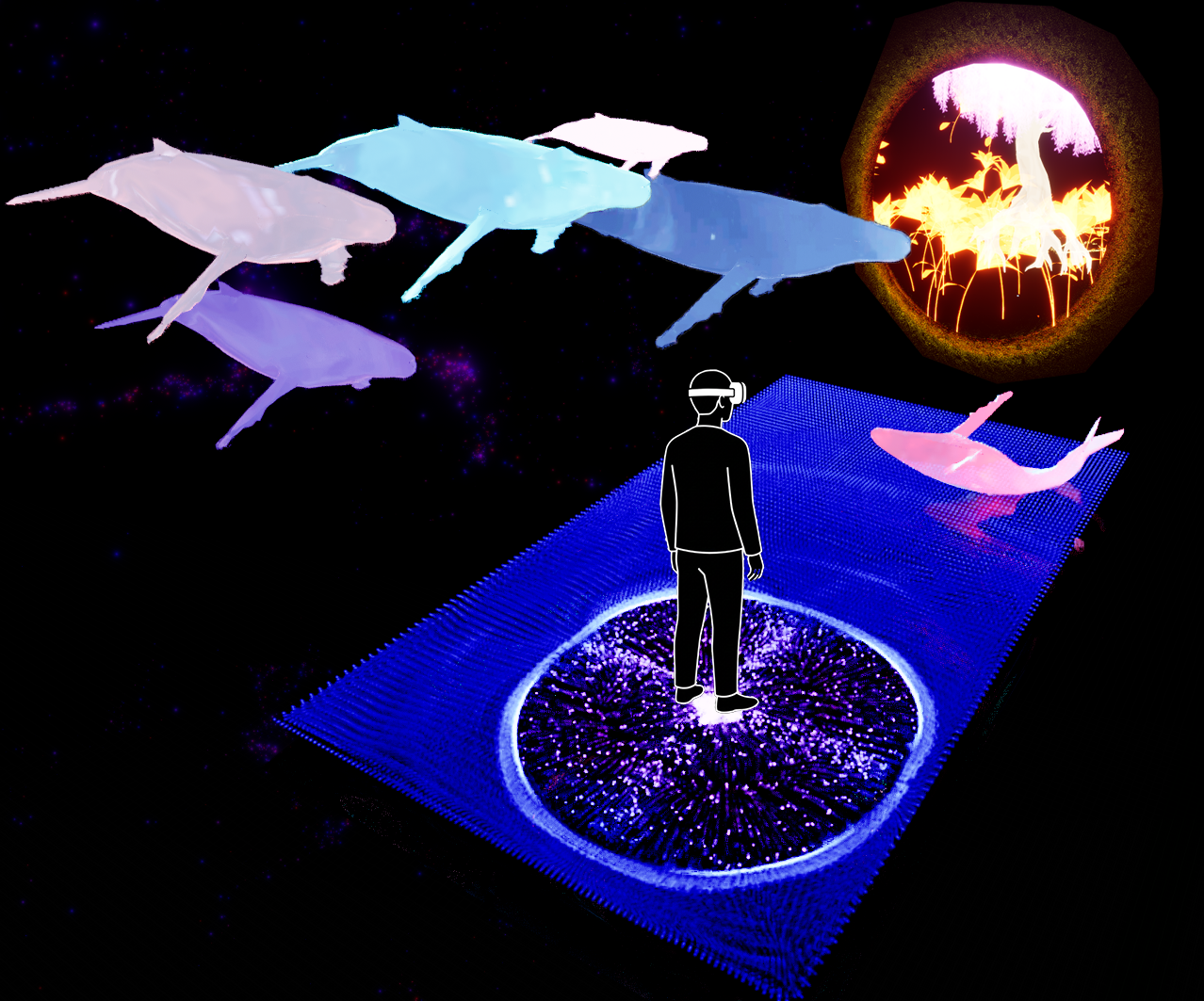}
    \caption{}
    \label{fig:scene3}
  \end{subfigure}
  \hfill
  \begin{subfigure}[t]{0.252\textwidth}
    \centering
    \includegraphics[width=\linewidth]{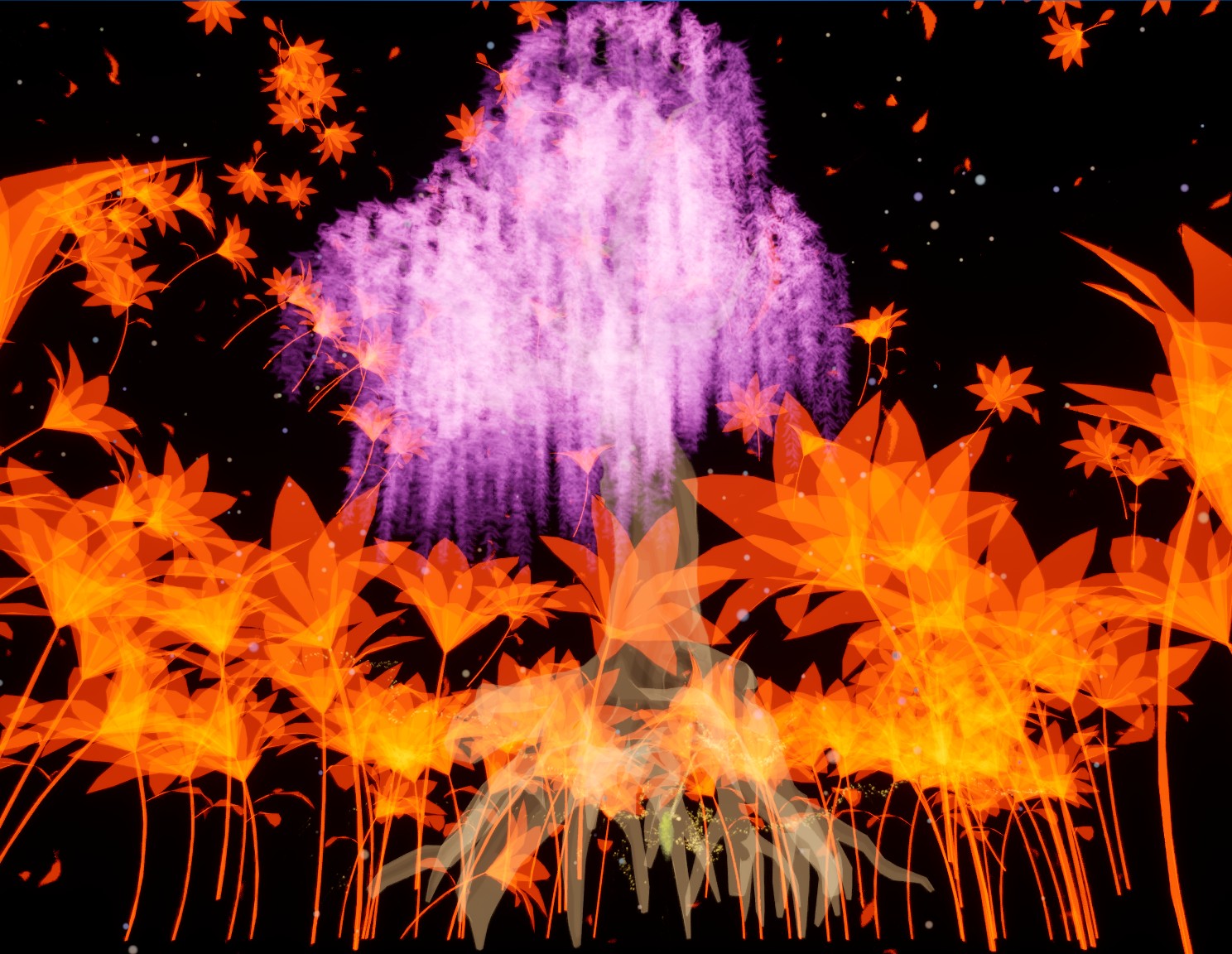}
    \caption{}
    \label{fig:scene4}
  \end{subfigure}

  \caption{(a) Scene~1 embeds a lightweight precision interaction with hit feedback. (b--c) Scenes~2--3 emphasize exploratory, atmosphere-driven interactions and continuity cues, including a portal that leads from AR to VR. (d) Scene~4 is a VR-only destination scene.}
  \label{fig:scenes_overview}
\end{figure*}

The experience comprises four sequential scenes. Scenes~1--3 combine projection with AR, and Scene~4 is a VR destination. Although exploration is open-ended, Scene~1 includes a lightweight precision interaction task to elicit reliable action--outcome feedback across conditions. Together, these scenes allow transition experience to be examined across both open-ended engagement and more controlled interaction.
\begin{itemize}[noitemsep, topsep=0pt]
  \item \textbf{Scene 1:} a bounded projection room (wall+floor) with rigid spheres (\cref{fig:scene1}). Participants push spheres toward randomly appearing targets with hit feedback, enabling repeated precision interactions and object-level transitions across projection and AR.
  \item \textbf{Scene 2:} an open hybrid space centered on a dynamic particle formation, augmented with animated skeletal and static elements (\cref{fig:scene2}). Interactions are primarily exploratory and atmosphere-driven.
  \item \textbf{Scene 3:} a larger projection--AR environment with large animated assets and an AR-to-VR portal (\cref{fig:scene3}), emphasizing navigation and continuity cues rather than precise targeting.
  \item \textbf{Scene 4:} a VR-only destination scene with flowers, trees, and roaming animals (\cref{fig:scene4}).
\end{itemize}

\subsubsection{Asset Selection and Placement}
To examine asset-type sensitivity (RQ3), we include three common real-time asset types and instantiate each with representative assets and dynamics:
\begin{itemize}[noitemsep, topsep=0pt]
  \item \textbf{Static meshes:} rigid geometry in Scene 1 (simple spheres) and Scene~2 (detail-rich meshes, e.g., \emph{flowers}).
  \item \textbf{Skeletal meshes:} animated assets with different motion scales and frequencies (Scene 2 \emph{butterflies} vs.\ Scene 3 \emph{whales}).
  \item \textbf{Particle systems:} a dynamic, spatially distributed particle \emph{cloud} in Scene 2 with diffuse boundaries and collective motion.
\end{itemize}
 
\subsubsection{Transition Placement}

\begin{figure}[t]
  \centering
  \begin{subfigure}[t]{0.49\linewidth}
    \centering
    \includegraphics[width=\linewidth]{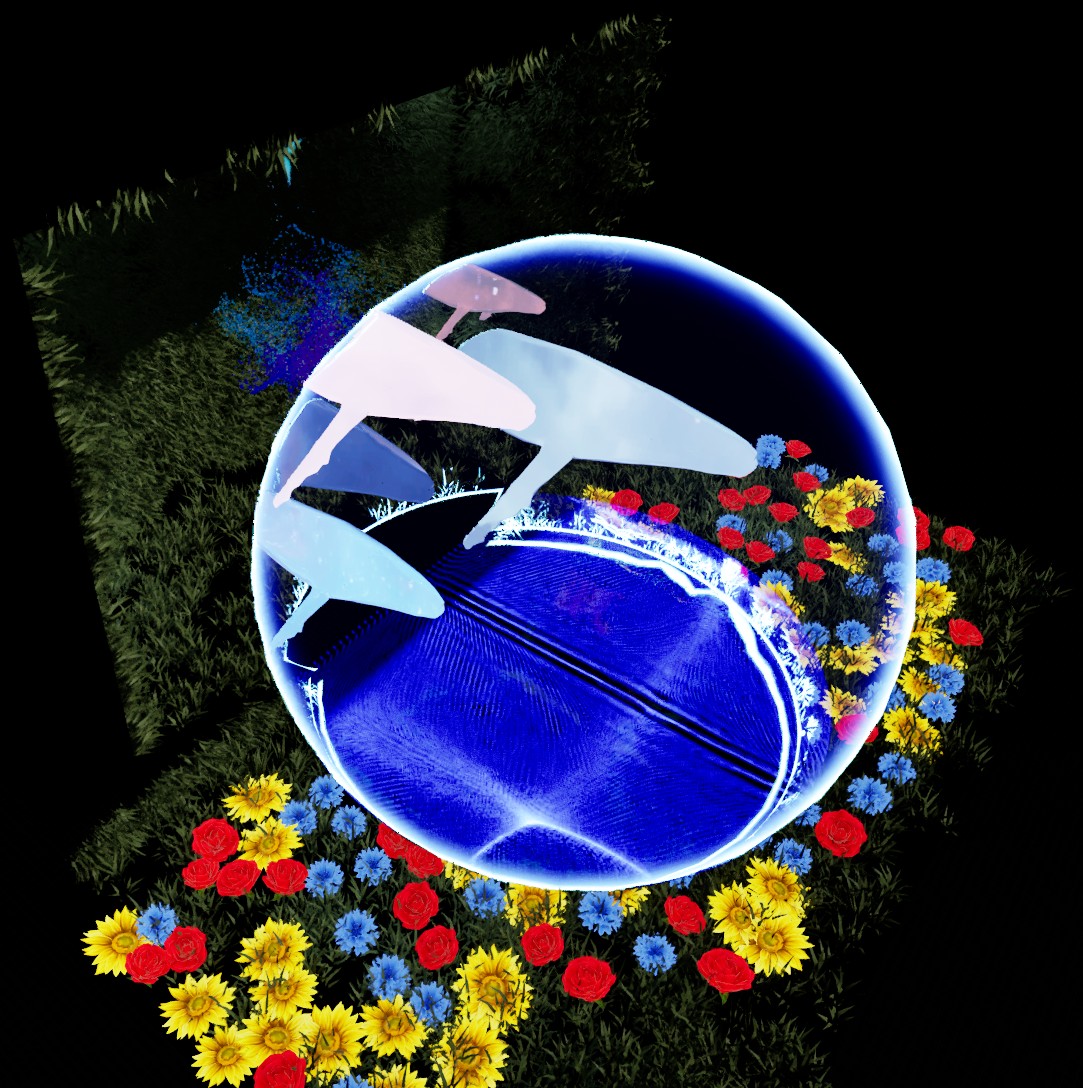}
    \caption{}
    \label{fig:Transition}
  \end{subfigure}
  \hfill
  \begin{subfigure}[t]{0.498\linewidth}
    \centering
    \includegraphics[width=\linewidth]{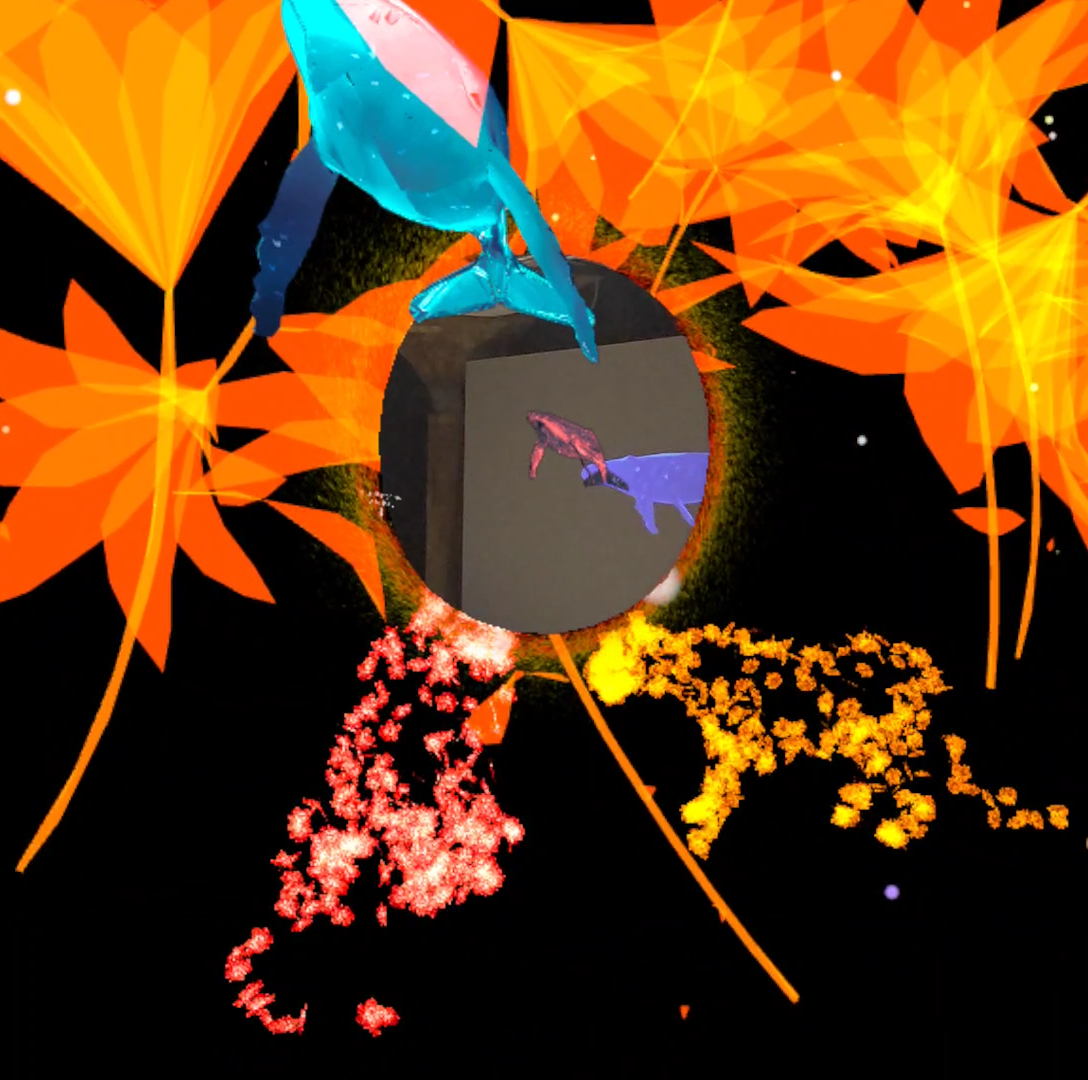}
    \caption{}
    \label{fig:portal}
  \end{subfigure}
  \caption{Scene-level transitions in the immersive art application. (a) A localized expansion effect gradually reveals the next projection--AR scene. (b) the AR--VR portal viewed from VR, showing animated assets spanning projection, AR, and the portal boundary to support continuity cues across modalities.}
  \label{fig:scene_transitions}
\end{figure}

Transitions are integrated into exploration rather than isolated tasks. In Scenes 1--3, object-level transitions occur as users move assets across projection boundaries, enabling bidirectional transfer between projection and AR. Scene-level transitions are designed to match the embodied and spatial nature of the immersive art experience. Between projection--AR scenes, participants trigger a localized expansion effect that gradually replaces the current scene and reveals the next while preserving spatial continuity (\cref{fig:Transition}). This effect was used as a spatially anchored scene replacement rather than as a standalone transition technique to be compared against alternatives. In Scene 3, a portal allows both users and selected animated assets to traverse between AR and VR, providing a previewable and physically traversable boundary between environments (\cref{fig:portal}). Together, these transition placements allow us to study continuity across both object-level projection--AR boundaries and scene-level projection--AR--VR changes within the same immersive art experience.

\subsection{Experimental Conditions Control}
\label{subsec:Control}
We compared two within-subject conditions while holding scenes, content, and interaction constant. The calibrated condition represents the intended operating state of the system. After baseline calibration and appearance tuning, we introduced the bundled inconsistency condition as a controlled diagnostic probe spanning spatial alignment, visual appearance, and cross-device latency. Its purpose was not to define acceptable calibration, color-matching, or latency thresholds, nor to provide hardware targets, but to create a noticeable contrast that made transition disruptions more perceptible and discussable for participants.

\paragraph{Spatial Alignment.}
We perturb the calibration-derived projection--HMD mapping (\cref{subsubsec:calibration}) by introducing position/rotation/scale offsets during proxy alignment, which globally shifts all cross-device object correspondences.

\paragraph{Visual Appearance.}
We applied uniform color grading changes to the projection rendering pipeline by reducing saturation and contrast, while leaving AR rendering unchanged. This created a controlled appearance mismatch without altering geometry or interaction.

\paragraph{Cross-Device Latency.}
We add a fixed delay to the synchronization channel between the projection instance and the standalone HMD instance (\cref{subsubsec:system}), increasing the temporal offset between corresponding updates.

\section{User Study}
We conducted a mixed-methods within-subjects study to examine how bundled transition inconsistencies and asset types influence perceived transition quality, presence, and workload in our immersive art application. Ethical approval for this study was obtained from the Research Ethics Committee of the School of Computer Science and Statistics, Trinity College Dublin.

\subsection{Participants}
We recruited 24 participants (7 female, 17 male; age 23--36, $M=27.7$, $SD=4.0$) from graduate students and the general public. All reported normal or corrected-to-normal vision and first-time exposure to the system. Prior VR experience varied (None: 4, Basic: 3, Intermediate: 10, Expert: 7).

\subsection{Apparatus}
Participants used the hybrid projection--MR system (\cref{subsubsec:system}, \cref{sec:Application}) in a single-user setup. The physical space, display layout, interaction devices, and rendering settings were identical across conditions.

\subsection{Procedure}
Participants provided informed consent and completed the Immersive Tendencies Questionnaire (ITQ) \cite{Witmer98}. They explored the four scenes in a fixed order with no time limits. During the lightweight precision interaction task in Scene~1, we also recorded participants' task completion time as a behavioral measure. We provided a brief interaction overview (controller trigger), participants could ask whether an element was interactive during the experiment, but received no information about conditions or expected outcomes.

The study comprised two sessions: a calibrated baseline and a bundled inconsistency condition (\cref{subsec:Control}). In the inconsistency condition, we applied a 10\,cm rightward offset, a $10^\circ$ clockwise rotation, and a 15\% scale increase to AR relative to projection; reduced projection saturation and contrast by 50\%; and added 100\,ms delay to synchronization. These magnitudes were chosen through iterative testing to produce a noticeable contrast while still allowing participants to complete the experience. They were fixed across participants to preserve comparability and support direct cross-session reflection. They should therefore be interpreted as experimental perturbations for a diagnostic contrast, rather than as perceptual thresholds, acceptable error bounds, or hardware targets. To mitigate potential order effects, participants were assigned to two counterbalanced groups: G1 experienced the calibrated condition followed by the inconsistent condition, whereas G2 experienced the inconsistent condition followed by the calibrated condition.

After each session, participants completed the Presence Questionnaire (PQ) \cite{Witmer98} and NASA-TLX \cite{TLX}. After both sessions, 21 participants completed a semi-structured interview on perceived transition cues and asset-type sensitivity, with prompts encouraging direct cross-session comparisons. The study lasted $\sim$40 minutes.

\subsection{Data Analysis}
PQ items were rated on a 7-point scale (0--6), and presence was computed as the mean of the 19 items. NASA-TLX was computed using Raw TLX by averaging the six subscales (0--100). For within-subject comparisons between sessions, we assessed normality of difference scores using the Shapiro--Wilk test \cite{Shapiro65}. When normality was met, we used paired-samples t-tests and report effect sizes as Cohen's $d_z$ \cite{Cohen88}. When normality was violated, we used Wilcoxon signed-rank tests and report effect sizes as $r=|z|/\sqrt{N}$. ITQ scores were summarized descriptively and used in exploratory correlations with outcomes.

For the lightweight precision interaction task in Scene 1, we recorded task completion time and compared conditions using the same within-subject procedure as for the questionnaire measures, first assessing normality of difference scores with the Shapiro–Wilk test and then applying a paired-samples t-test when normality was met.

Interview transcripts were analyzed using inductive thematic analysis. The first author conducted iterative open coding across all transcripts, identifying segments related to perceived transition cues, experiential consequences, and asset-type sensitivity. Codes were grouped into candidate themes and refined through repeated comparison across participants and conditions. Emerging themes and their interpretations were discussed with the co-authors in multiple review sessions to improve consistency and resolve ambiguities. As the analysis was exploratory and inductive, we did not compute formal intercoder reliability; instead, we emphasize iterative theme refinement and collaborative interpretation.

\section{Results}
\subsection{Quantitative Results}

\Cref{fig:paired_results} shows within-subject changes between the calibrated and bundled inconsistency conditions. In our study design, the quantitative comparison was not intended to make the obvious point that degradation worsens experience, but to confirm that the bundled inconsistency condition produced an intended diagnostic contrast. The changes in presence, workload, and Scene~1 task performance therefore establish that participants experienced the two conditions differently. Because the manipulation deliberately combined spatial, appearance, and latency discrepancies, these statistics are not intended to isolate cue-specific effects. The main interpretive contribution comes from the following qualitative analysis, which examines which cues participants noticed, how they attributed their effects, and how cue salience varied across interaction contexts and asset types.

\subsubsection{Presence}
Presence decreased under the inconsistency condition ($M=3.85$, $SD=0.70$) compared to the calibrated condition ($M=4.82$, $SD=0.53$) (\cref{fig:paired_pq}). Because the difference scores were not normally distributed (Shapiro--Wilk $p=.002$), we report a Wilcoxon signed-rank test as the primary analysis. Presence was significantly lower in the inconsistency condition (Median $=3.92$, IQR $[3.46, 4.26]$) than in the calibrated condition (Median $=4.71$, IQR $[4.41, 5.20]$), $W=0.00$, $p<.001$, $z=-4.29$, $r=.88$.

\subsubsection{Task Load}
NASA-TLX increased under the inconsistency condition ($M=36.32$, $SD=15.85$) compared to the calibrated condition ($M=23.47$, $SD=11.19$) (\cref{fig:paired_tlx}). Difference scores were approximately normally distributed (Shapiro--Wilk $p=.525$). A paired-samples t-test showed a significant increase ($t(23)=4.33$, $p<.001$; $d_z=0.88$), with a mean paired difference of $\Delta M=12.85$ and a 95\% CI of $[6.71,\,18.99]$. 

\subsubsection{Task completion time in Scene~1}
\begin{figure*}[t]
  \centering
  \begin{subfigure}[t]{0.32\textwidth}
    \centering
    \includegraphics[width=\linewidth]{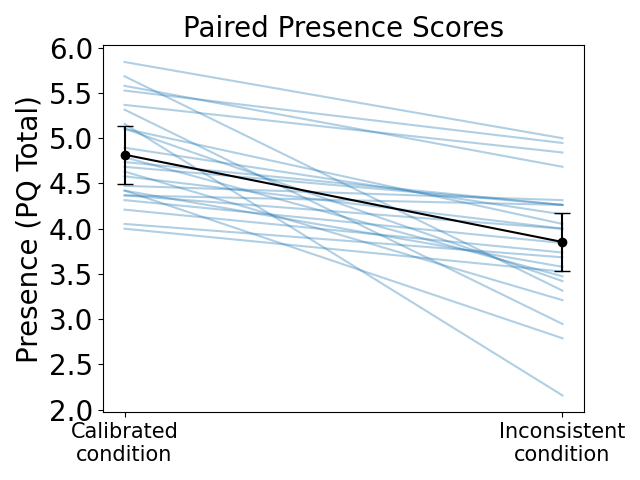}
    \caption{Presence}
    \label{fig:paired_pq}
  \end{subfigure}
  \hfill
  \begin{subfigure}[t]{0.32\textwidth}
    \centering
    \includegraphics[width=\linewidth]{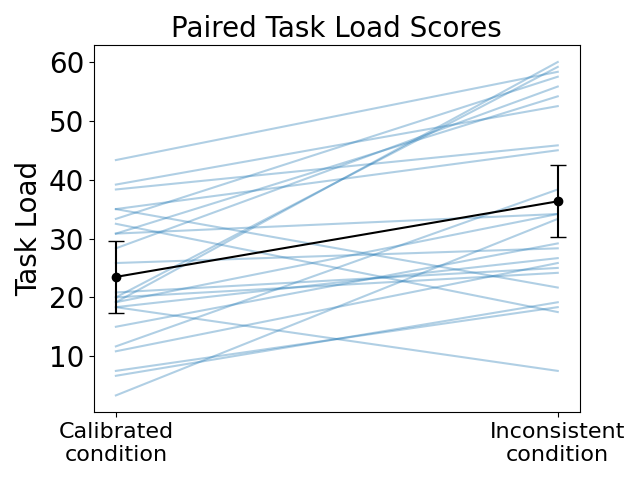}
    \caption{Task load}
    \label{fig:paired_tlx}
  \end{subfigure}
  \hfill
  \begin{subfigure}[t]{0.32\textwidth}
    \centering
    \includegraphics[width=\linewidth]{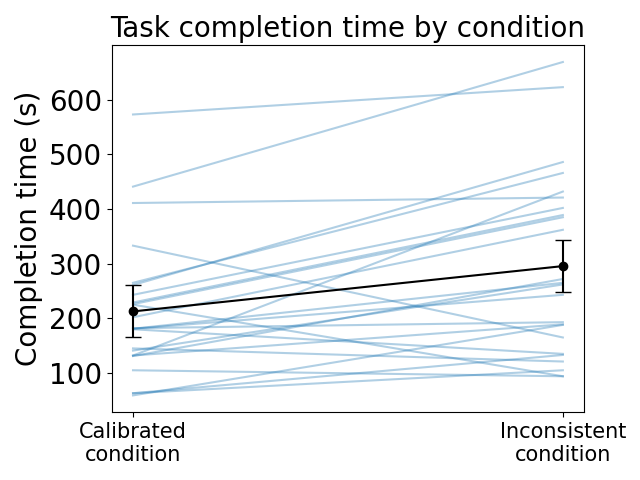}
    \caption{Task completion time}
    \label{fig:Time}
  \end{subfigure}
  \caption{Paired plots comparing the calibrated and bundled inconsistency conditions. Each line represents one participant. Black markers indicate the mean; error bars show 95\% CIs. (a) Presence, (b) task load, and (c) Scene~1 task completion time.}
  \label{fig:paired_results}
\end{figure*}

In Scene~1, participants completed the lightweight precision interaction task faster in the calibrated condition ($M = 212.67$\,s, $SD = 124.40$\,s) than in the bundled inconsistency condition ($M = 295.58$\,s, $SD = 166.15$\,s) (\cref{fig:Time}). Difference scores were approximately normally distributed (Shapiro--Wilk $p = .892$).  A paired-samples t-test showed a significant increase ($t(23) = 3.60$, $p = .0015$; $d_z = 0.74$), with a mean increase of $82.92$\,s (95\% CI [$35.29$, $130.54$]).

\subsubsection{Immersive Tendencies}
Participants reported moderate immersive tendencies (ITQ: $M=4.02$, $SD=0.69$). Exploratory correlations suggested that higher ITQ scores were associated with higher presence in the calibrated condition and with larger presence reductions under inconsistencies, no reliable associations were observed for NASA-TLX. We report these analyses as exploratory. We did not apply corrections for multiple comparisons.

\subsection{Qualitative Findings}
We interviewed 21 of 24 participants after both sessions. Themes are organized around perceived transition cues (RQ1), asset-type sensitivity (RQ3), and continuity beyond individual factors.

\subsubsection{Perceived transition cues (RQ1, RQ2)}
\textbf{Spatial alignment} was frequently reported (13/21) and described holistically as \emph{``offset''} or \emph{``not aligned''}. Misalignment was most salient during precision interactions (spheres; 12/21),  where accurate aiming and collision outcomes were required. In these situations, even small discrepancies in alignment reduced confidence in interaction accuracy. One participant, for example, stated that \emph{``I wasn’t sure where the ball would actually go after I pushed it''} (P07), indicating that spatial misalignment disrupted action--outcome mapping during task-oriented interactions.

\textbf{Visual appearance} mismatch was the most commonly noticed difference (20/21), described as \emph{``reduced vividness''}, \emph{``weaker contrast''}, or \emph{``a less saturated look''}. It was especially salient on visually rich static assets such as \emph{flowers} (9/21), whose detailed surface appearance made cross-display differences easier to notice. At the same time, several participants framed color inconsistency as primarily aesthetic rather than functionally disruptive. For example, one participant stated that \emph{``Color has the least impact.''} (P01), suggesting that appearance mismatches were salient perceptually but less likely to interfere with interaction or task performance.

\textbf{Temporal responsiveness} (latency) was mentioned by 10/21 participants as \emph{``lag''} or \emph{``delay''}. When mentioned, latency was often associated with reduced responsiveness during interaction, which participants linked to lower perceived control and increased frustration. For example, one participant stated that \emph{``Second session was ... lagging and rendering at 5 fps, which was annoying''} (P12). Notably, a small number of participants attributed such delays to system or network performance rather than to transition design itself, yet still described a negative impact on overall experience. Taken together, these accounts suggest that latency operated as a salient feedback cue that could undermine interaction quality and perceived continuity, even when participants did not explicitly frame it as a transition inconsistency.

\subsubsection{Asset-type sensitivity (RQ3)}
\textbf{Static meshes} were most frequently cited as revealing transition inconsistencies. In our coding, 12 participants identified the sphere assets as making spatial misalignment especially salient during precision interactions such as aiming and pushing, where small deviations directly affected perceived interaction accuracy. In addition, 9 participants noted that visually rich static meshes, such as the flower elements, made color differences more apparent due to their detailed surface appearance and saturation. Overall, participants described static meshes as providing clear and stable visual reference cues, which made discrepancies between projection and MR representations easier to detect, and these assets were often used as benchmarks for judging whether transitions felt consistent across sessions.

\textbf{Skeletal meshes} elicited mixed but generally weaker sensitivity to cross-session inconsistencies than static assets. Eight participants reported that animated butterflies and whales felt largely similar during transfer; for example, one participant noted that \emph{``the butterflies were basically the same, and the whales in the two sessions felt very similar''} (P22). When differences were mentioned, participants tended to describe them holistically (e.g., realism or overall impression) rather than attributing them to a single factor. One participant felt that the butterfly transition in the second session was \emph{``not as realistic as in the first session''} and speculated that both offset and color might contribute (P23). Another participant noted a size change, stating that \emph{``the size of the butterfly in the second session was quite big compared to the first session''} (P14). Overall, these accounts suggest that animated assets were often evaluated in terms of overall motion impression during transfer, and their continuous animation could reduce attention to small cross-session discrepancies that were readily exposed by rigid static geometry.

\textbf{Particle effects} were perceived as most tolerant. They were frequently described as less sensitive to cross-session inconsistencies than rigid geometry. Many participants reported that particle-based visuals remained convincing even when other assets exhibited noticeable differences. At the same time, a small number of participants noticed subtle changes in how particle structure was perceived during transfer. For example, one participant described a shift from a \emph{``3D feeling to a 2D feeling''} when dragging the particle system from AR back into the projection view (P05). Overall, participants tended to evaluate particle effects in terms of overall impression and atmosphere, and their distributed visual structure often reduced the salience of small geometric or appearance mismatches.

\subsubsection{Continuity beyond individual cues}
\textbf{Object identity} was generally preserved: most participants reported that they recognized the projection-based and MR-based representations as the same object during transitions, even when discrepancies were present. A subset of participants (n = 7) explicitly emphasized that contextual cues, in particular interaction feedback and continuous visual transformation during transition, helped maintain this identity judgment. One participant highlighted the role of agency, noting that \emph{``because I was the person in control ... my interaction ... helped me identify''} (P21). Another participant referred to the continuous transformation during transfer, stating that \emph{``as it crossed the screen, its change was continuous''} (P23). In addition, a subset of participants (n = 5) reported that after understanding the interaction logic and adapting to the systematic offset, they continued to treat the object as identical across modalities even when misalignment became more salient during transfer, stating that \emph{``when I understand what I'm doing, it's no longer a problem''} (P23).

\textbf{Scene transitions} were discussed in terms of experiential smoothness and staging, including the spherical expansion between projection--AR scenes and the portal-based transition between AR and VR. Seven participants explicitly praised the portal as a memorable mechanism; one participant noted \emph{``I like a lot the portal part, that was working really well''} (P18). The expansion-based scene transition was commonly described as a gradual transformation of the surrounding space, for example as \emph{``a feeling of the entire space gradually switching to a whale scene''} (P06). When comparing the two sessions, 8 participants reported noticing differences in the expansion-based transition, primarily in terms of perceived smoothness or latency. In contrast, spatial misalignment and color mismatch were rarely mentioned at the scene level, and none of the participants explicitly attributed scene-level differences to these factors. Some participants additionally suggested that the connective moment around the portal could be made more vivid or expressive (P22), indicating that scene transitions were evaluated not only for continuity but also for narrative and experiential framing.

\textbf{CR connectedness} varied across participants. When reflecting on how projection, AR, and VR related as a whole, participants reported both connected and fragmented impressions depending on how salient inconsistencies became across sessions. Six participants described the overall experience as coherent, attributing connectedness to narrative flow and persistent entities traversing modalities (e.g., \emph{``when I was in the VR world, I did turn around and I could still see the whales moving through the portal... that element of moving from one to the other made it feel very connected''} (P21)). Conversely, five participants reported a more fragmented impression in the less consistent session; one participant commented that the second session produced \emph{``It feels a bit disjointed, because subconsciously you still know they are two separate devices.''} (P10). Overall, connectedness depended on whether continuity cues dominated attention or whether system-level discrepancies became foregrounded.

\section{Discussion}
\subsection{Interpretation of Findings}

Our interpretation does not rest on the fact that the inconsistent condition reduced experience quality. Instead, we use the calibrated--inconsistent comparison to examine which transition cues became salient, how participants attributed their effects, and how those effects varied across interaction contexts and asset types.

\subsubsection{RQ1: Cues shaping perceived CR transition quality}
Guided by our transitional-affordance lens (\cref{subsec:Factors}), we examined cues that support perceived continuity across projection and the HMD. Interview findings aligned with this framing: participants most frequently cited spatial misalignment and appearance mismatch (color) as salient cross-display inconsistencies, while latency was described as a feedback-related signal that reduced perceived responsiveness and control. Participants typically articulated these cues holistically (e.g., \emph{``offset''}, \emph{``less vivid''}, \emph{``lag''}) rather than decomposing individual geometric parameters, suggesting that transition quality emerges from integrated impressions of cross-modal consistency.

\subsubsection{RQ2: How do these cues affect users' immersive experience?}

Quantitative results showed that the bundled inconsistency condition substantially reduced presence, increased workload, and slowed performance in the Scene~1 precision interaction task. Because the manipulation bundled spatial, appearance, and latency discrepancies, the quantitative measures alone do not isolate the contribution of each cue. We therefore use the qualitative accounts to interpret how participants perceived and attributed these disruptions across interaction contexts.

Qualitative accounts suggest two complementary mechanisms. First, participants most often connected spatial inconsistency and latency to action-facing aspects of the experience. These cues undermined users' sense of control and predictability during interaction, increasing uncertainty about where objects would be and how they would respond. Their effects were most salient during task-like interactions that required precise action--outcome mapping, such as aiming and collision. This helps explain why the bundled inconsistency condition was associated with slower Scene~1 performance and higher workload: participants had to invest more effort to interpret and compensate for unreliable cross-display feedback.

Second, appearance mismatch shaped the experiential and aesthetic tone of the scenes. Participants frequently noticed reduced vividness, weaker contrast, and lower visual richness, but often framed these differences as aesthetic degradation rather than direct barriers to successful interaction. In this sense, appearance mismatch may have contributed to the reduction in presence by weakening visual involvement and aesthetic coherence, even though participants did not typically describe it as impairing task performance.

This distinction helps position our findings against prior work on transition context and transition frictions. Although prior work has primarily examined environment-level transitions or switching between reality contexts, it provides a useful lens for interpreting how transition disruptions become consequential in our object-level setting. Feld et al. show that task context can shift users' preferences toward efficient transition techniques when repeated switching is required~\cite{Feld24}. Our findings point to a related lower-level requirement: when an object transition is embedded in a precision interaction, users may judge coherence less by the visibility of the transition effect itself and more by whether spatial and temporal feedback preserves a reliable action--outcome mapping.

At the same time, our exploratory and scene-level moments show that transition quality in immersive art is not reducible to efficiency or task performance. Participants also evaluated transitions through vividness, smoothness, aesthetic coherence, and connectedness. This resonates with qualitative accounts of CR frictions, where transition problems can emerge when the system disrupts users' spatial mental model or makes realities feel insufficiently connected~\cite{vonWillich25}. In our case, these broader frictions appeared when projection and HMD content felt less coherent as one experience, and when scene transitions were judged through smoothness, pacing, and connectedness rather than task efficiency alone.

Overall, these findings suggest that participants experienced the bundled inconsistencies through context-dependent mechanisms rather than through a single uniform pathway: spatial and temporal discrepancies were described as most consequential for perceived control during interaction, while appearance mismatch was mainly described as shaping visual involvement and aesthetic coherence.

\subsubsection{RQ3: Asset-type sensitivity}
Interview findings indicate that perceived sensitivity to inconsistencies differed by asset type. We interpret this pattern as reflecting differences in the perceptual reference cues provided by each asset structure during transitions. Static meshes were most often described as diagnostic of inconsistencies. Their rigid geometry and stable surface appearance provided clear spatial and visual anchors, making offsets and color mismatches easier to detect, especially during precision interactions. Skeletal meshes elicited weaker and more mixed sensitivity. Several participants reported that animated butterflies and whales felt similar across sessions, and continuous motion and deformation appeared to draw attention toward overall animation impression rather than small cross-display discrepancies. Particle effects were frequently perceived as least affected. Participants often evaluated them in terms of atmosphere and overall impression, and their distributed structure and lack of stable boundaries tended to reduce the salience of small spatial or appearance mismatches, although a small number of participants still noticed changes in perceived depth or dimensionality during the transition. These observations position our findings in relation to prior object-level CR transition work. Prior studies have emphasized the importance of preserving object continuity across devices, realities, or 2D/3D representation spaces~\cite{Fischer23,Wang22_2,Liao25,Lee22,Schwajda23}. Our findings are consistent with this emphasis, but add that continuity is not experienced uniformly across object content. Stable geometry and detailed surface appearance can provide useful reference cues for object identity, yet the same cues also make spatial and appearance mismatches more visible. Conversely, articulated motion and distributed visual structure can shift attention toward motion impression or atmosphere, making small cross-display discrepancies less salient. The three asset categories in our study are not intended as an exhaustive taxonomy of all CR objects, but as a practical lens for reasoning about how recurring structural properties shape the salience of transition inconsistencies.

\subsection{Implications for CR Design}
Our findings suggest several practical implications for designing experience-driven CR transitions in hybrid environments. Although our implementation combined projection and an MR HMD, we expect these insights to also inform hybrid environments that integrate other forms of 2D displays with HMDs, insofar as they involve cross-display continuity and transition coherence.

\paragraph{Allocate consistency budgets according to interaction and asset demands.}
The transitioning object should not be treated as a uniform unit of analysis: consistency requirements appear to depend on the asset's structural cues, interaction context, and experiential role. Spatial alignment should be treated as a first-order requirement when users need reliable action--outcome mapping, such as aiming, pushing, or judging collisions. In these cases, rigid or detail-rich assets can provide strong reference cues for object identity and interaction feedback, but the same cues also make spatial or appearance mismatches more visible. Designers should therefore allocate stricter consistency budgets to diagnostic assets and interaction-critical moments, such as static meshes used for precision targeting or collision feedback.

When perfect alignment cannot be guaranteed, systems may benefit from reducing reliance on exact geometric correspondence, for example by avoiding precision targeting across display boundaries or by introducing interaction techniques that tolerate small offsets. Designers can also use assets whose structure is less likely to expose small discrepancies. Animated or distributed effects often reduced the salience of minor inconsistencies in our study, suggesting that they may be useful when the asset's role is primarily atmospheric or when perfect alignment is difficult. More broadly, dynamic or distributed visuals, such as particles, haze, or fluid-like effects, can act as perceptual ``buffers'' during transitions by masking small spatial or appearance mismatches.

\paragraph{Maintain appearance consistency to support aesthetic coherence.}
Color differences were highly noticeable and shaped the experiential tone of scenes, even when they were not always described as functionally disruptive. Maintaining consistent color appearance across projection and HMD rendering can strengthen aesthetic coherence, particularly for assets with rich surface detail where mismatches become more salient. When exact color matching is difficult (e.g., due to projector or display characteristics), designers may intentionally adopt stylized or reduced-detail renderings that lessen the visibility of cross-device appearance discrepancies.

\paragraph{Treat latency as a feedback cue that shapes perceived control.}
Participants frequently interpreted temporal mismatch as reduced responsiveness, which undermined perceived control and increased frustration. This suggests that cross-display latency should be minimized particularly during moments where users directly manipulate objects or expect immediate feedback. When latency is unavoidable, designers may consider mitigation strategies such as smoothing, prediction, or transition animations that mask small temporal offsets without compromising users' sense of agency. At the system level, this also implies optimizing the communication pipeline used for cross-device state synchronization, so that updates are transmitted and applied as consistently and efficiently as possible.

\paragraph{Make continuity cues explicit across objects and scenes.}
Designers can strengthen CR coherence by making continuity cues explicit, including continuous visual transformations, stable interaction feedback, and legible cross-display trajectories. Persistent entities that traverse projection, AR, and VR and continue to exist across transitions can further serve as anchors that reinforce both object identity and scene-level connectedness. This resonates with prior qualitative CR work showing that users rely on anchor points and spatial mental models to connect realities~\cite{vonWillich25}, while highlighting how, in an experience-driven art context, such anchors can also carry narrative continuity. 

\paragraph{Stage scene transitions according to whether the goal is efficiency or experiential continuity.}
Prior work suggests that efficient cuts or fades can be preferable for frequent, task-driven switches, whereas portals and other visible transitions can support preview, pre-orientation, presence, and continuity when environments differ substantially or transitions are less frequent~\cite{Feld24,Husung19,Pointecker22,Gottsacker26}. Our findings suggest that experience-driven immersive art often resembles the latter case: participants evaluated scene changes through smoothness, pacing, and expressive staging, and several participants explicitly described the portal as memorable or effective. The expansion transition was also described as the surrounding space gradually switching, suggesting that staged visual change can make scene transition part of the artwork rather than only a navigation mechanism. Designers could therefore use portals, expansion effects, and persistent entities when the transition itself should carry spatial or narrative continuity, while reserving shorter and less visible techniques for repeated task-oriented switching.

\paragraph{Preserve or release prior context depending on the experience goal.}
Prior work on spatial cognitive residue shows that transitions can influence how strongly users disengage from one virtual environment before entering the next~\cite{Gottsacker24}. This adds an important qualification to continuity-oriented design: coherence is not always achieved by preserving as much of the previous context as possible. In unrelated task-switching contexts, designers may want transitions to help users release prior spatial assumptions and redirect attention to the new environment. In narrative or exhibition contexts, however, selected forms of residue may be desirable. In our study, both the AR--VR portal and the expansion transition appeared to support continuity by preserving selected cues across the transition. The portal maintained continuity through preview, physical traversal, and persistent animated entities, while the expansion transition was often interpreted as a gradual transformation of the surrounding space. We interpret this cautiously, because the study was not designed to isolate these elements causally. Nevertheless, these observations suggest that preserving selected cues, such as motion direction, atmosphere, and narrative momentum, can help scene transitions function as part of a continuous artwork rather than as complete resets.

\paragraph{Leverage complementary device roles rather than replacing one modality with another.}
Hybrid projection--MR systems are valuable not because projection or HMDs replace one another, but because they can distribute experiential roles across devices. Projection can be driven by a desktop PC to render complex and high-fidelity environmental scenes on a large physical surface. Prior work has shown that large projection displays and CAVE-like immersive environments can, in some contexts, support comparable or even stronger immersion, realism, presence, engagement, and spatial perception than HMD-based VR~\cite{Theo21,Theo23,Juan09,Wischgoll24,Tcha17}. Meanwhile,  optical see-through (OST) and video see-through (VST) HMDs, as well as emerging MR glasses, allow users to perceive the surrounding physical environment and projected content while also supporting viewpoint-dependent augmentation, embodied 3D interaction, and flexible object manipulation beyond the physical display surface. In this sense, projection can provide the environmental background and shared spatial context, while the HMD can provide interactive, personalized, and spatially registered virtual content.

This complementarity helps explain why cross-display transition coherence is central to our findings: when projection and HMD content are combined, spatial, appearance, and temporal inconsistencies can undermine the very continuity that makes the hybrid setup valuable. We therefore do not argue that hybrid projection--MR is inherently superior to HMD-only or projection-only systems. Rather, it is a promising design direction for contexts where the additional setup effort is justified by the need to combine large-scale environmental immersion with flexible MR interaction. Future work should directly compare these configurations to evaluate their trade-offs in task performance, presence, comfort, spatial perception, and user preference.

\section{Limitations and Future Work}
This study has several limitations and directions for future work. First, our participant sample was male-skewed, with 17 male and 7 female participants. Future studies should recruit more gender-balanced and demographically diverse samples to better capture a wider range of user perspectives.

Second, we manipulated spatial alignment, visual appearance, and latency cues as a bundled inconsistency condition. This design created a realistic and noticeable diagnostic contrast, but prevents attributing the observed effects to any single cue. Accordingly, the results should not be read as specifying acceptable calibration, color-matching, or latency thresholds for projection--MR systems, nor as hardware performance targets. Instead, they characterize how a deliberately noticeable bundle of inconsistencies shaped participants’ perceptions and reflections in this particular immersive art context. Future studies could vary these cues independently to estimate thresholds, interactions, and context-dependent tolerance ranges. 

Third, findings are grounded in one hybrid setup and one experience-driven application; replication across hardware, display layouts, interaction techniques, and task demands is needed for generalizability. Several hardware-related constraints also affected our implementation. Even in the calibrated condition, the PC rendering and synchronization pipeline introduced perceptible delay between the projected content and the HMD, so the baseline was not completely latency-free. Our setup also used only two projection planes, wall and floor, which may provide a lower level of immersion than larger exhibition-scale environments. Calibration quality was further limited by HMD hardware and tracking stability: as participants moved, the HMD-based calibrator could drift slightly, leading to small but visible spatial misalignments over time. In addition, although we tuned both the PC-side projected content and projector image settings before the study to approximate the appearance of the corresponding MR-rendered content, the Quest 3 passthrough still introduced residual blur, reduced sharpness, exposure variation, and color differences that could not be fully eliminated. These residual passthrough artifacts were held constant across conditions because the same HMD, projection setup, room lighting, and projection settings were used throughout the study. Future work could improve the technical baseline by reducing rendering and synchronization latency, scaling the setup to larger and more immersive projection configurations, improving tracking and calibration robustness, investigating systematic color calibration and passthrough-aware compensation, and exploring OST MR devices to reduce passthrough-induced blur and color distortion in projection--MR experiences.

Fourth, we studied a single-user configuration. Multi-user exhibitions often require a fixed projection viewpoint without head-coupled parallax, which may alter perceived alignment and continuity and introduce social effects on transition experience. Future work could therefore examine multi-user and collaborative projection--MR settings, including how shared projection content, individual HMD views, turn-taking, co-located interaction, and group attention affect transition coherence and perceived connectedness across realities.

Finally, the experience was primarily exploratory and qualitative insights relied on post-session self-reports; future work could add controlled tasks, behavioral measures such as accuracy and error rates, and repeated-session studies to examine whether users adapt to transition inconsistencies over time.

\section{Conclusion}
We presented an experience-driven investigation of CR transitions using an immersive art application built on a hybrid projection--MR system. Guided by a transitional affordance lens, we examined three cue dimensions for transition coherence: spatial alignment, visual appearance consistency, and cross-display latency. Rather than treating the bundled inconsistency condition as a threshold test or hardware target, we used it as a diagnostic contrast to reveal how transition-relevant disruptions are perceived and interpreted.

Our mixed-methods results show that projection--MR transition quality depends on how specific cues become salient across interaction contexts and asset types. Spatial misalignment primarily disrupted action--outcome predictability, appearance mismatch affected aesthetic coherence, and latency reduced perceived responsiveness. These effects were further shaped by asset structure: rigid and detail-rich static assets exposed mismatches more clearly, whereas animated skeletal assets and distributed particle effects often shifted attention toward motion impression or atmosphere.

Together, these findings suggest that coherent hybrid projection--MR experiences require both overall technical consistency and an understanding of where specific inconsistencies become most perceptually and experientially consequential. They motivate asset-aware transition strategies that prioritize spatial and temporal consistency during interaction-critical moments, maintain appearance coherence for visually diagnostic assets, and support continuity across projection, AR, and VR. We release the core system as an open-source Unreal Engine plugin to support future projection--MR CR research and prototyping.

\section*{Supplemental Material pointers}
\label{sec:supplemental_materials}
Supplemental videos are available at \url{https://osf.io/rqmct}. The open-source Unreal Engine plugin \textit{HUICRSync}, together with its documentation and example
project, is publicly available on GitHub at \url{https://github.com/XiangpengFu/HUICRSync-UnrealProject}. The version used in this work is archived on Zenodo at \url{https://doi.org/10.5281/zenodo.21380424}.

\acknowledgments{%
	This work was conducted with the financial support of the Research Ireland Centre for Research Training in Digitally-Enhanced Reality (d-real) under Grant No. 18/CRT/6224. For the purpose of Open Access, the author has applied a CC BY public copyright licence to any Author Accepted Manuscript version arising from this submission.%
}

\bibliographystyle{abbrv-doi-hyperref}

\bibliography{template}

@inproceedings{Feiner91,
author = {Feiner, Steven and Shamash, Ari},
title = {Hybrid user interfaces: breeding virtually bigger interfaces for physically smaller computers},
year = {1991},
isbn = {0897914511},
publisher = {Association for Computing Machinery},
address = {New York, NY, USA},
url = {https://doi.org/10.1145/120782.120783},
doi = {10.1145/120782.120783},
booktitle = {Proceedings of the 4th Annual ACM Symposium on User Interface Software and Technology},
pages = {9–17},
numpages = {9},
location = {Hilton Head, South Carolina, USA},
series = {UIST '91}
}

@INPROCEEDINGS{Satkowski23,
  author={Satkowski, Marc and Méndez, Julián},
  booktitle={2023 IEEE International Symposium on Mixed and Augmented Reality Adjunct (ISMAR-Adjunct)}, 
  title={Fantastic Hybrid User Interfaces and How to Define Them}, 
  year={2023},
  volume={},
  number={},
  pages={247-250},
  doi={10.1109/ISMAR-Adjunct60411.2023.00057}}

@ARTICLE{Reipschlager21,
  author={Reipschlager, Patrick and Flemisch, Tamara and Dachselt, Raimund},
  journal={IEEE Transactions on Visualization and Computer Graphics}, 
  title={Personal Augmented Reality for Information Visualization on Large Interactive Displays}, 
  year={2021},
  volume={27},
  number={2},
  pages={1182-1192},
  doi={10.1109/TVCG.2020.3030460}}

@article{Zagermann22,
author = {Zagermann, Johannes and Hubenschmid, Sebastian and Balestrucci, Priscilla and Feuchtner, Tiare and Mayer, Sven and Ernst, Marc and Schmidt, Albrecht and Reiterer, Harald},
year = {2022},
month = {09},
pages = {},
title = {Complementary interfaces for visual computing},
volume = {64},
journal = {it - Information Technology},
doi = {10.1515/itit-2022-0031}
}

@INPROCEEDINGS{Liang23,
  author={Liang, Hai-Ning and Yu, Lingyun and Liarokapis, Fotis},
  booktitle={2023 IEEE Conference on Virtual Reality and 3D User Interfaces Abstracts and Workshops (VRW)}, 
  title={Workshop: Mixing Realities: Cross-Reality Visualization, Interaction, and Collaboration}, 
  year={2023},
  volume={},
  number={},
  pages={298-300},
  doi={10.1109/VRW58643.2023.00069}}

@article{Milgram94,
author = {Milgram, Paul and Takemura, Haruo and Utsumi, Akira and Kishino, Fumio},
year = {1994},
month = {01},
pages = {},
title = {Augmented reality: A class of displays on the reality-virtuality continuum},
volume = {2351},
journal = {Telemanipulator and Telepresence Technologies},
doi = {10.1117/12.197321}
}

@inproceedings{Pavavimol25,
author = {Pavavimol, Tippayaporn and Ometov, Aleksandr and Valkama, Mikko and Thibault, Mattia},
title = {Transitions between Realities: A Systematic Review on the Usage of {XR} Systems for Bridging Reality and Virtuality},
year = {2025},
isbn = {9798400713910},
publisher = {Association for Computing Machinery},
address = {New York, NY, USA},
url = {https://doi.org/10.1145/3706370.3727858},
doi = {10.1145/3706370.3727858},
booktitle = {Proceedings of the 2025 ACM International Conference on Interactive Media Experiences},
pages = {158–174},
numpages = {17},
location = {
},
series = {IMX '25}
}

@inproceedings{Butscher18,
author = {Butscher, Simon and Hubenschmid, Sebastian and M\"{u}ller, Jens and Fuchs, Johannes and Reiterer, Harald},
title = {Clusters, Trends, and Outliers: How Immersive Technologies Can Facilitate the Collaborative Analysis of Multidimensional Data},
year = {2018},
isbn = {9781450356206},
publisher = {Association for Computing Machinery},
address = {New York, NY, USA},
url = {https://doi.org/10.1145/3173574.3173664},
doi = {10.1145/3173574.3173664},
pages = {1–12},
numpages = {12},
location = {Montreal QC, Canada},
series = {CHI '18}
}

@article{Zhao25,
author = {Zhao, Lixiang and Isenberg, Tobias and Xie, Fuqi and Liang, Hai-Ning and Yu, Lingyun},
title = {{SpatialTouch}: Exploring Spatial Data Visualizations in Cross-Reality},
year = {2025},
issue_date = {Jan. 2025},
publisher = {IEEE Educational Activities Department},
address = {USA},
volume = {31},
number = {1},
issn = {1077-2626},
url = {https://doi.org/10.1109/TVCG.2024.3456368},
doi = {10.1109/TVCG.2024.3456368},
journal = {IEEE Transactions on Visualization and Computer Graphics},
month = jan,
pages = {897–907},
numpages = {11}
}

@ARTICLE{Liao25,
  author={Liao, Shuqi and Chaudhri, Sparsh and Karwa, Maanas K. and Popescu, Voicu},
  journal={IEEE Transactions on Visualization and Computer Graphics}, 
  title={Seamless{VR}: Bridging the Immersive to Non-Immersive Visualization Divide}, 
  year={2025},
  volume={31},
  number={5},
  pages={2806-2816},
  doi={10.1109/TVCG.2025.3549564}}

@INPROCEEDINGS{Cavallo19,
  author={Cavallo, Marco and Dholakia, Mishal and Havlena, Matous and Ocheltree, Kenneth and Podlaseck, Mark},
  booktitle={2019 IEEE Conference on Virtual Reality and 3D User Interfaces (VR)}, 
  title={Dataspace: A Reconfigurable Hybrid Reality Environment for Collaborative Information Analysis}, 
  year={2019},
  volume={},
  number={},
  pages={145-153},
  doi={10.1109/VR.2019.8797733}}

@inproceedings{Reipschlager19,
author = {Reipschl\"{a}ger, Patrick and Dachselt, Raimund},
title = {Design{AR}: Immersive 3D-Modeling Combining Augmented Reality with Interactive Displays},
year = {2019},
isbn = {9781450368919},
publisher = {Association for Computing Machinery},
address = {New York, NY, USA},
url = {https://doi.org/10.1145/3343055.3359718},
doi = {10.1145/3343055.3359718},
booktitle = {Proceedings of the 2019 ACM International Conference on Interactive Surfaces and Spaces},
pages = {29–41},
numpages = {13},
location = {Daejeon, Republic of Korea},
series = {ISS '19}
}

@article{Yu23,
author = {Deng, Yu  and Wang, Yao},
year = {2023},
month = {02},
pages = {012048},
title = {Research on holographic display and technology application of art museum based on immersive design},
volume = {2425},
journal = {Journal of Physics: Conference Series},
doi = {10.1088/1742-6596/2425/1/012048}
}

@inproceedings{Langner21,
author = {Langner, Ricardo and Satkowski, Marc and B\"{u}schel, Wolfgang and Dachselt, Raimund},
title = {{MARVIS}: Combining Mobile Devices and Augmented Reality for Visual Data Analysis},
year = {2021},
isbn = {9781450380966},
publisher = {Association for Computing Machinery},
address = {New York, NY, USA},
url = {https://doi.org/10.1145/3411764.3445593},
doi = {10.1145/3411764.3445593},
articleno = {468},
numpages = {17},
location = {Yokohama, Japan},
series = {CHI '21}
}

@inproceedings{Fischer23,
author = {Fischer, Robin and Lian, Wei-Xiang and Wang, Shiann-Jang and Hsu, Wei-En and Fu, Li-Chen},
title = {Seamless Virtual Object Transitions: Enhancing User Experience in Cross-Device Augmented Reality Environments},
year = {2023},
isbn = {978-3-031-43400-6},
publisher = {Springer-Verlag},
address = {Berlin, Heidelberg},
url = {https://doi.org/10.1007/978-3-031-43401-3_26},
doi = {10.1007/978-3-031-43401-3_26},
booktitle = {Extended Reality: International Conference, XR Salento 2023, Lecce, Italy, September 6-9, 2023, Proceedings, Part I},
pages = {397–409},
numpages = {13},
location = {Lecce, Italy}
}

@inproceedings{Lee22,
author = {Lee, Benjamin and Cordeil, Maxime and Prouzeau, Arnaud and Jenny, Bernhard and Dwyer, Tim},
title = {A Design Space For Data Visualisation Transformations Between 2D And 3D In Mixed-Reality Environments},
year = {2022},
isbn = {9781450391573},
publisher = {Association for Computing Machinery},
address = {New York, NY, USA},
url = {https://doi.org/10.1145/3491102.3501859},
doi = {10.1145/3491102.3501859},
booktitle = {Proceedings of the 2022 CHI Conference on Human Factors in Computing Systems},
articleno = {25},
numpages = {14},
location = {New Orleans, LA, USA},
series = {CHI '22}
}

@inproceedings{Husung19,
author = {Husung, Malte and Langbehn, Eike},
title = {Of Portals and Orbs: An Evaluation of Scene Transition Techniques for Virtual Reality},
year = {2019},
isbn = {9781450371988},
publisher = {Association for Computing Machinery},
address = {New York, NY, USA},
url = {https://doi.org/10.1145/3340764.3340779},
doi = {10.1145/3340764.3340779},
booktitle = {Proceedings of Mensch Und Computer 2019},
pages = {245–254},
numpages = {10},
location = {Hamburg, Germany},
series = {MuC '19}
}

@INPROCEEDINGS{Pointecker22,
  author={Pointecker, Fabian and Friedl, Judith and Schwajda, Daniel and Jetter, Hans-Christian and Anthes, Christoph},
  booktitle={2022 IEEE International Symposium on Mixed and Augmented Reality (ISMAR)}, 
  title={Bridging the Gap Across Realities: Visual Transitions Between Virtual and Augmented Reality}, 
  year={2022},
  volume={},
  number={},
  pages={827-836},
  doi={10.1109/ISMAR55827.2022.00101}}

@inproceedings{Pointecker24,
author = {Pointecker, Fabian and Friedl-Knirsch, Judith and Jetter, Hans-Christian and Anthes, Christoph},
title = {From Real to Virtual: Exploring Replica-Enhanced Environment Transitions along the Reality-Virtuality Continuum},
year = {2024},
isbn = {9798400703300},
publisher = {Association for Computing Machinery},
address = {New York, NY, USA},
url = {https://doi.org/10.1145/3613904.3642844},
doi = {10.1145/3613904.3642844},
booktitle = {Proceedings of the 2024 CHI Conference on Human Factors in Computing Systems},
articleno = {799},
numpages = {13},
location = {Honolulu, HI, USA},
series = {CHI '24}
}

@ARTICLE{Feld24,
  author={Feld, Nico and Bimberg, Pauline and Weyers, Benjamin and Zielasko, Daniel},
  journal={IEEE Transactions on Visualization and Computer Graphics}, 
  title={Simple and Efficient? Evaluation of Transitions for Task-Driven Cross-Reality Experiences}, 
  year={2024},
  volume={30},
  number={12},
  pages={7601-7618},
  doi={10.1109/TVCG.2024.3356949}}

@article{Rhee13,
  title={Affordance in Interactive Media Art Exhibition},
  author={Rhee On Jeong and Seungho Park},
  journal={International Journal of Asia Digital Art and Design},
  volume={17},
  number={3},
  pages={93-99},
  year={2013},
  doi={10.20668/adada.17.3_93}
}

@INPROCEEDINGS{Cools22,
  author={Cools, Robbe and Gottsacker, Matt and Simeone, Adalberto and Bruder, Gerd and Welch, Greg and Feiner, Steven},
  booktitle={2022 IEEE International Symposium on Mixed and Augmented Reality Adjunct (ISMAR-Adjunct)}, 
  title={Towards a Desktop-AR Prototyping Framework: Prototyping Cross-Reality Between Desktops and Augmented Reality}, 
  year={2022},
  volume={},
  number={},
  pages={175-182},
  doi={10.1109/ISMAR-Adjunct57072.2022.00040}
}

@ARTICLE{Cools25,
  author={Cools, Robbe and Maerevoet, Inne and Gottsacker, Matt and Simeone, Adalberto L.},
  journal={IEEE Transactions on Visualization and Computer Graphics}, 
  title={Comparison of Cross-Reality Transition Techniques between 3D and 2D Display Spaces in Desktop–AR Systems}, 
  year={2025},
  volume={},
  number={},
  pages={1-8},
  doi={10.1109/TVCG.2025.3549907}
}

@inproceedings{Rau25,
author = {Rau, Tobias and Isenberg, Tobias and Koehn, Andreas and Sedlmair, Michael and Lee, Benjamin},
title = {Traversing Dual Realities: Investigating Techniques for Transitioning 3D Objects between Desktop and Augmented Reality Environments},
year = {2025},
isbn = {9798400713941},
publisher = {Association for Computing Machinery},
address = {New York, NY, USA},
url = {https://doi.org/10.1145/3706598.3713949},
doi = {10.1145/3706598.3713949},
booktitle = {Proceedings of the 2025 CHI Conference on Human Factors in Computing Systems},
articleno = {1236},
numpages = {16},
location = {
},
series = {CHI '25}
}

@ARTICLE{Witmer98,
  author={Witmer, Bob G. and Singer, Michael J.},
  journal={Presence}, 
  title={Measuring Presence in Virtual Environments: A Presence Questionnaire}, 
  year={1998},
  volume={7},
  number={3},
  pages={225-240},
  doi={10.1162/105474698565686}}

@article{Shapiro65,
 ISSN = {00063444, 14643510},
 URL = {http://www.jstor.org/stable/2333709},
 author = {S. S. Shapiro and M. B. Wilk},
 journal = {Biometrika},
 number = {3/4},
 pages = {591--611},
 publisher = {[Oxford University Press, Biometrika Trust]},
 title = {An Analysis of Variance Test for Normality (Complete Samples)},
 urldate = {2026-01-08},
 volume = {52},
 year = {1965},
 doi = {10.2307/2333709}
}

@book{Cohen88,
  author    = {Cohen, Jacob},
  title     = {Statistical Power Analysis for the Behavioral Sciences},
  edition   = {2nd},
  year      = {1988},
  publisher = {Lawrence Erlbaum Associates},
  address   = {Hillsdale, NJ},
  doi={10.4324/9780203771587}
}

@misc{vangogh26,
  author       = {{Exhibition Hub}},
  title        = {{Van Gogh Exhibition: The Immersive Experience}},
  year         = {2026},
  howpublished = {\url{https://vangoghexpo.com/}},
  note         = {Accessed: 2026-01-09}
}

@ARTICLE{Cools25_DP,
  author={Cools, Robbe and Han, Jihae and Esteves, Augusto and Simeone, Adalberto L.},
  journal={IEEE Transactions on Visualization and Computer Graphics}, 
  title={From Display to Interaction: Design Patterns for Cross-Reality Systems}, 
  year={2025},
  volume={31},
  number={5},
  pages={3129-3139},
  doi={10.1109/TVCG.2025.3549893}}

@inproceedings{Wang22,
author = {Wang, Nanjia and Maurer, Frank},
title = {A Design Space for Single-User Cross-Reality Applications},
year = {2022},
isbn = {9781450397193},
publisher = {Association for Computing Machinery},
address = {New York, NY, USA},
url = {https://doi.org/10.1145/3531073.3531116},
doi = {10.1145/3531073.3531116},
booktitle = {Proceedings of the 2022 International Conference on Advanced Visual Interfaces},
articleno = {31},
numpages = {5},
location = {Frascati, Rome, Italy},
series = {AVI '22}
}

@ARTICLE{Schwajda23,
    
AUTHOR={Schwajda, Daniel  and Friedl, Judith  and Pointecker, Fabian  and Jetter, Hans-Christian  and Anthes, Christoph },
           
TITLE={Transforming graph data visualisations from 2D displays into augmented reality 3D space: A quantitative study},
          
JOURNAL={Frontiers in Virtual Reality},
          
VOLUME={Volume 4 - 2023},
  
YEAR={2023},
  
URL={https://www.frontiersin.org/journals/virtual-reality/articles/10.3389/frvir.2023.1155628},
  
DOI={10.3389/frvir.2023.1155628},
  
ISSN={2673-4192},
}

@article{KAVE18,
author = {Gon\c{c}alves, Afonso and Berm\'{u}dez, Sergi},
title = {KAVE: Building Kinect Based {CAVE} Automatic Virtual Environments, Methods for Surround-Screen Projection Management, Motion Parallax and Full-Body Interaction Support},
year = {2018},
issue_date = {June 2018},
publisher = {Association for Computing Machinery},
address = {New York, NY, USA},
volume = {2},
number = {EICS},
url = {https://doi.org/10.1145/3229092},
doi = {10.1145/3229092},
journal = {Proc. ACM Hum.-Comput. Interact.},
month = jun,
articleno = {10},
numpages = {15}
}

@article{TLX,
author = {Sandra G. Hart},
title ={Nasa-Task Load Index ({NASA-TLX}); 20 Years Later},
journal = {Proceedings of the Human Factors and Ergonomics Society Annual Meeting},
volume = {50},
number = {9},
pages = {904-908},
year = {2006},
doi = {10.1177/154193120605000909},
URL = {https://doi.org/10.1177/154193120605000909
},
eprint = {https://doi.org/10.1177/154193120605000909
}
}

@INPROCEEDINGS{Wang22_2,
  author={Wang, Nanjia and Maurer, Frank},
  booktitle={2022 IEEE International Symposium on Mixed and Augmented Reality Adjunct (ISMAR-Adjunct)}, 
  title={User-Centered Prototyping for Single-User Cross-Reality Virtual Object Transitions}, 
  year={2022},
  volume={},
  number={},
  pages={171-174},
  doi={10.1109/ISMAR-Adjunct57072.2022.00039}}

@inproceedings{Gottsacker26,
author = {Gottsacker, Matt and Hmaiti, Yahya and Maslych, Mykola and Furuya, Hiroshi and DeGuzman, Jasmine Joyce and Bruder, Gerd and Welch, Gregory F. and LaViola, Joseph J.},
title = {From One World to Another: Interfaces for Efficiently Transitioning Between Virtual Environments},
year = {2026},
isbn = {9798400722783},
publisher = {Association for Computing Machinery},
address = {New York, NY, USA},
url = {https://doi.org/10.1145/3772318.3791912},
doi = {10.1145/3772318.3791912},
booktitle = {Proceedings of the 2026 CHI Conference on Human Factors in Computing Systems},
articleno = {1690},
numpages = {17},
location = {
},
series = {CHI '26}
}

@INPROCEEDINGS{Gottsacker24,
  author={Gottsacker, Matt and Furuya, Hiroshi and Battistel, Laura and Jimenez, Carlos Pinto and LaMontagna, Nicholas and Bruder, Gerd and Welch, Gregory F.},
  booktitle={2024 IEEE International Symposium on Mixed and Augmented Reality (ISMAR)}, 
  title={Exploring Spatial Cognitive Residue and Methods to Clear Users’ Minds When Transitioning Between Virtual Environments}, 
  year={2024},
  volume={},
  number={},
  pages={1000-1009},
  doi={10.1109/ISMAR62088.2024.00116}}

@inproceedings{vonWillich25,
author = {von Willich, Julius and Nelles, Frank and Tseng, Wen-Jie and Gugenheimer, Jan and G\"{u}nther, Sebastian and M\"{u}hlh\"{a}user, Max},
title = {A Qualitative Investigation of User Transitions and Frictions in Cross-Reality Applications},
year = {2025},
isbn = {9798400713941},
publisher = {Association for Computing Machinery},
address = {New York, NY, USA},
url = {https://doi.org/10.1145/3706598.3713921},
doi = {10.1145/3706598.3713921},
booktitle = {Proceedings of the 2025 CHI Conference on Human Factors in Computing Systems},
articleno = {808},
numpages = {18},
location = {
},
series = {CHI '25}
}

@article{Theo23,
  title={{CAVE} and {HMD}: distance perception comparative study},
  author={Combe, Th{\'e}o and Chardonnet, Jean-R{\'e}my and Merienne, Fr{\'e}d{\'e}ric and Ovtcharova, Jivka},
  journal={Virtual Reality},
  volume={27},
  number={3},
  pages={2003--2013},
  year={2023},
  doi={10.1007/s10055-023-00787-y},
  publisher={Springer}
}

@INPROCEEDINGS{Theo21,
  author={Combe, Théo and Chardonnet, Jean-Rémy and Merienne, Frédéric and Ovtcharova, Jivka},
  booktitle={2021 IEEE Conference on Virtual Reality and 3D User Interfaces Abstracts and Workshops (VRW)}, 
  title={{CAVE} vs. {HMD} in Distance Perception}, 
  year={2021},
  volume={},
  number={},
  pages={448-449},
  doi={10.1109/VRW52623.2021.00106}}

@article{Juan09,
    author = {Juan, M. Carmen and Pérez, David},
    title = {Comparison of the Levels of Presence and Anxiety in an Acrophobic Environment Viewed via {HMD} or {CAVE}},
    journal = {Presence: Teleoperators and Virtual Environments},
    volume = {18},
    number = {3},
    pages = {232-248},
    year = {2009},
    month = {06},
    doi = {10.1162/pres.18.3.232},
    url = {https://doi.org/10.1162/pres.18.3.232},
    eprint = {https://direct.mit.edu/pvar/article-pdf/18/3/232/1624916/pres.18.3.232.pdf},
}

@INPROCEEDINGS{Wischgoll24,
  author={Wischgoll, Thomas},
  booktitle={2024 IEEE Conference on Virtual Reality and 3D User Interfaces Abstracts and Workshops (VRW)}, 
  title={Toward the Comparison of Different VR Devices for Visualization}, 
  year={2024},
  volume={},
  number={},
  pages={520-524},
  doi={10.1109/VRW62533.2024.00100}}

@inproceedings{Tcha17,
author = {Tcha-Tokey, Katy and Loup-Escande, Emilie and Christmann, Olivier and Richir, Simon},
title = {Effects on User Experience in an Edutainment Virtual Environment: Comparison Between {CAVE} and {HMD}},
year = {2017},
isbn = {9781450352567},
publisher = {Association for Computing Machinery},
address = {New York, NY, USA},
url = {https://doi.org/10.1145/3121283.3121284},
doi = {10.1145/3121283.3121284},
booktitle = {Proceedings of the European Conference on Cognitive Ergonomics},
pages = {1–8},
numpages = {8},
location = {Ume\r{a}, Sweden},
series = {ECCE '17}
}

\end{document}